\documentclass[a4paper,11pt]{article}

\usepackage{jcappub} 

\usepackage[T1]{fontenc} 
\usepackage{subfigure}
\usepackage[Symbol]{upgreek}

\usepackage{graphicx}
\usepackage{caption}
\usepackage{xcolor}
\usepackage[normalem]{ulem}
\usepackage[capitalize]{cleveref}
\usepackage{lscape}

\usepackage{float}
\usepackage{array}
\newcolumntype{P}[1]{>{\centering\arraybackslash}p{#1}}
\newcolumntype{C}[1]{>{\centering\arraybackslash}m{#1}}

\usepackage{tcolorbox}

\usepackage{tikz}
\usepackage[compat=1.1.0]{tikz-feynman}

\usepackage{siunitx}

\DeclareSIUnit \s {\second}
\DeclareSIUnit \ns {\nano\second}
\DeclareSIUnit \mus {\micro\second}
\DeclareSIUnit \ms {\milli\second}
\DeclareSIUnit \MB {\mega\byte}
\DeclareSIUnit \GB {\giga\byte}
\DeclareSIUnit \TB {\tera\byte}
\DeclareSIUnit \PB {\peta\byte}
\DeclareSIUnit \Mbps {\mega\bit/\s}
\DeclareSIUnit \Gbps {\giga\bit/\s}
\DeclareSIUnit \Tbps {\tera\bit/\s}
\DeclareSIUnit \Pbps {\peta\bit/\s}
\DeclareSIUnit \kton {\kilo\tonne} 
\DeclareSIUnit \kt {\kilo\tonne}
\DeclareSIUnit \Mt {\mega\tonne}
\DeclareSIUnit \eV {\electronvolt}
\DeclareSIUnit \keV {\kilo\electronvolt}
\DeclareSIUnit \MeV {\mega\electronvolt}
\DeclareSIUnit \GeV {\giga\electronvolt}
\DeclareSIUnit \TeV {\tera\electronvolt}
\DeclareSIUnit \PeV {\peta\electronvolt}
\DeclareSIUnit \EeV {\exa\electronvolt}
\DeclareSIUnit \m {\meter}
\DeclareSIUnit \cm {\centi\meter}
\DeclareSIUnit \in {\inchcommand}
\DeclareSIUnit \km {\kilo\meter}
\DeclareSIUnit \kV {\kilo\volt}   
\DeclareSIUnit \kW {\kilo\watt}
\DeclareSIUnit \MW {\mega\watt}
\DeclareSIUnit \MHz {\mega\hertz}
\DeclareSIUnit \mrad {\milli\radian}
\DeclareSIUnit \year {years}
\DeclareSIUnit \POT {POT}
\DeclareSIUnit \sig {$\sigma$}
\DeclareSIUnit\parsec{pc}
\DeclareSIUnit\kpc{kpc}
\DeclareSIUnit\Gpc{Gpc}
\DeclareSIUnit\lightyear{ly}
\DeclareSIUnit\foot{ft}
\DeclareSIUnit\ft{ft}
\DeclareSIUnit \ppb{ppb}
\DeclareSIUnit \ppt{ppt}
\DeclareSIUnit \samples{S}
\DeclareSIUnit \pe{PE}

\newcommand{\enu}{\E_\enu}

\title{\boldmath Multi-messenger High-Energy Signatures of Decaying Dark Matter and the Effect of Background Light}

\author[a,1]{B. Skrzypek,\note{Corresponding author.}}
\author[b,c]{M. Chianese,}
\author[a]{C. A. Arg\"uelles}

\affiliation[a]{Department of Physics \& Laboratory for Particle Physics and Cosmology,\\ Harvard University, Cambridge, MA 02138, USA}
\affiliation[b]{Dipartimento di Fisica ``Ettore Pancini'', Università degli studi di Napoli ``Federico II'', Complesso Univ. Monte S. Angelo, I-80126 Napoli, Italy}
\affiliation[c]{INFN - Sezione di Napoli, Complesso Univ. Monte S. Angelo, I-80126 Napoli, Italy}

\emailAdd{bskrzypek@g.harvard.edu}
\emailAdd{chianese@na.infn.it}
\emailAdd{carguelles@fas.harvard.edu}

\abstract{
The IceCube Neutrino Observatory at the South Pole has measured astrophysical neutrinos using through-going and starting events in the TeV to PeV energy range.
The origin of these astrophysical neutrinos is still largely unresolved, and among their potential sources could be dark matter decay.
Measurements of the astrophysical flux using muon neutrinos are in slight tension with starting event measurements.
This tension is driven by an excess observed in the energy range of 40-200 TeV with respect to the through-going expectation.
Previous works have considered the possibility that this excess may be due to heavy dark matter decay and have placed constraints using gamma-ray and neutrino data.
However, these constraints are not without caveats since they rely on the modeling of the astrophysical neutrino flux and the sources of gamma-ray emission.
In this work, we derive background-agnostic galactic and extragalactic constraints on decaying dark matter by considering Tibet AS$_\gamma$ data, Fermi-LAT diffuse data, and the IceCube high-energy starting event sample.
For the gamma-ray limits, we investigate the uncertainties on secondary emission from electromagnetic cascades during propagation arising from the unknown intensity of the extragalactic background light.
We find that such uncertainties amount to a variation of up to $\sim 55\%$ in the gamma-ray limits derived with extragalactic data.
Our results imply that a significant fraction of the astrophysical neutrino flux could be due to dark matter and that ruling it out depends on the assumptions on the gamma-ray and neutrino background.
The latter depends on the yet unidentified sources.}

\begin{document}
\maketitle
\flushbottom

\section{Introduction \label{sec:intro}}

The measurement of High-Energy Starting Event (HESE) neutrinos in the TeV-PeV range by the IceCube Neutrino Observatory has introduced new questions in connection to their origin and properties~\cite{Ackermann:2019ows,Ackermann:2019cxh,Ackermann:2022rqc}. 
These open questions have the possibility of being answered through multi-messenger efforts in astroparticle physics~\cite{Meszaros:2019xej,AlvesBatista:2021gzc}.
In particular, the excess observed in the 7.5-year HESE sample, with its sensitivity to the Galactic center and, generally, to the entire sky, cannot originate entirely from the atmospheric neutrino background, nor does it lend itself to a simple astrophysical power-law distribution given the observations in other channels~\cite{IceCube:2020wum}.
Instead, a two-component power-law spectrum, the nature of which is in question, is favored by the data.
Moreover, the modeling of this astrophysical contribution is yet to be determined.
However, it is well-known that the production mechanisms involved in giving rise to highly energetic astrophysical neutrinos, namely the decay of charged pions originating from hadronic~\cite{Loeb:2006tw,Ambrosone:2020evo} and photohadronic~\cite{Winter:2013cla,Fiorillo:2021hty} interactions, also produce coincident gamma-rays at the source.
Most notably, this coincident emission was confirmed by the multi-messenger observation of the flaring blazar TXS 0506+056~\cite{MAGIC:2018sak,IceCube:2018dnn,IceCube:2018cha}.

However, the high level of gamma-ray suppression from astrophysical sources along with tensions among current IceCube datasets in measurements of the astrophysical spectrum make it difficult to characterize the nature of the source of these ultra-high-energy neutrinos.
With the assumption of a single unbroken power-law flux, the 10-year through-going (TG) muon sample, which is sensitive to neutrinos coming only from the Northern sky and with energies larger than $\SI{15}\TeV$, measures a spectral index of $2.37^{+0.09}_{-0.09}$~\cite{IceCube:2021uhz}.
This measurement is consistent with theoretical predictions of neutrinos originating from photohadronic processes~\cite{Waxman_1998} and with measurements of TXS 0506+056.
On the other hand, HESE, with sensitivity to the entire sky and to energies starting from $\SI{20}\TeV$, obtains a spectral index $2.87^{+0.20}_{-0.19}$~\cite{IceCube:2020wum}.
This suggests that there could be an additional component contributing to the neutrino flux that HESE observes that dominates at lower energies and comes from~\cite{Chen:2014gxa,IceCube:2015gsk,Gaggero:2015xza,Chianese:2016opp,Palladino:2016zoe,Vincent:2016nut,Palladino:2016xsy,Anchordoqui:2016ewn,Denton:2017csz,Chianese:2017jfa,Palladino:2018evm,Sui:2018bbh}.

Given this observation, heavy dark matter has been proposed as a potential source of the neutrinos in the HESE sample~\cite{Feldstein:2013kka,Esmaili:2013gha,Bai:2013nga,Bhattacharya:2014vwa,Esmaili:2014rma,Kachelriess:2018rty,Denton:2018aml,Bhattacharya:2019ucd,Chianese:2019kyl,Dekker:2019gpe} (see Ref.s~\cite{IceCube:2018tkk,Arguelles:2019boy,IceCube:2021sog} for the IceCube analyses).
Heavy decaying dark matter also gives rise to gamma-rays~\cite{Ciafaloni:2010ti,Buch:2015iya,Bauer:2020jay} which have been searched with gamma-ray telescopes.
Diffuse gamma-ray data places strong constraints on the dark matter lifetime for various decay channels, and these largely disfavor the dark matter hypothesis for IceCube's observations~\cite{Murase:2012xs,Murase:2015gea,Esmaili:2015xpa,Kalashev:2016cre,Cohen:2016uyg,HAWC:2017udy,Blanco:2018esa,Ishiwata:2019aet,Esmaili:2021yaw,Maity:2021umk,Chianese:2021jke}.
In order to properly constrain the size of the dark matter contribution in HESE, we need to establish the robustness of these gamma-ray bounds.

In this work, we elaborate on previous dark matter decay results by considering the impact that the Extragalactic Background Light (EBL) could have on the predicted gamma-ray spectrum from dark matter.
The presence of different EBL models is due to the fact that direct measurements are difficult and involve many uncertainties and tensions among observations, such as the disagreement between near-IR data and observations of very high-energy photons from extragalactic sources.
As a result, there are various techniques used to model the EBL~\cite{Dominguez:2010bv,Stecker:2016fsg,Finke:2009xi,Kneiske:2010pt,Gilmore2012,Helgason:2012xj,Inoue:2012bk,Franceschini2017,Andrews:2017ima}, some of which are based on local galaxy properties and others which are phenomenological approaches targeting a larger redshift range.
We simulate the propagation of photons, electrons, and positrons using three different models of the EBL distribution, and we compute the resulting gamma-ray spectra at Earth predicted by dark matter decay.
From there, we compare the fluxes to Tibet AS$_\gamma$ data~\cite{TibetASgamma:2021tpz} and Fermi-LAT diffuse measurements of the isotropic diffuse gamma-ray background (IGRB)~\cite{Fermi-LAT:2014ryh}, and we obtain lower bounds on the dark matter lifetime among the different EBL models that could have significant implications for existing limits.

The paper is organized as follows.
In~\Cref{sec:dm}, we discuss the high-energy neutrino and gamma-ray signatures of dark matter decay.
In~\Cref{sec:ebl}, we study the impact of the EBL uncertainties in the propagation of gamma-rays.
In~\Cref{sec:analysis}, we describe our analyses on neutrino and gamma-ray data sets and present our results.
Finally, in~\Cref{sec:conclusion} we conclude.

\section{Heavy Dark Matter Decay and its Signatures\label{sec:dm}}

Generally, models of heavy dark matter allow for decay modes where the final products are Standard Model particle-antiparticle pairs. 
In all the decay modes, neutrinos and gamma-rays are directly produced through the electroweak and hadronic cascades which immediately follow the decay as well as the electroweak radiative corrections which become relevant for dark matter masses larger than $\SI{100}\GeV$~\cite{Ciafaloni:2010ti,Bauer:2020jay}.
Together, these determine the neutrino and gamma-ray differential energy spectra, which we denote as ${\rm d}N_i/{\rm d}E$ with $i=\nu, \gamma$. This is used as the first input to compute the observed fluxes at Earth. 

Because of their nature, neutrinos injected by dark matter decay reach the Earth unperturbed, barring effects on their final state flavor configuration due to neutrino oscillations.
Therefore, the neutrino flux at the Earth can be easily calculated from the initial neutrino differential spectrum.
We can characterize dark matter sources as either galactic (G) or extragalactic (EG) in nature, and thus the final differential neutrino flux can be broken down into galactic and extragalactic contributions, namely
\begin{equation} \label{eq:nu_flux_tot}
    \frac{{\rm d}\phi_{\nu}}{{\rm d}E_\nu {\rm d}\Omega} = \frac{{\rm d}\phi^{\rm G}_{\nu}}{{\rm d}E_\nu {\rm d}\Omega} + \frac{{\rm d}\phi^{\rm EG}_{\nu}}{{\rm d}E_\nu {\rm d}\Omega}\,,
\end{equation}
where $E_\nu$ denotes the neutrino energy and $\Omega$ denotes the solid angle.

For the galactic contribution, we assume a generalized Navarro-Frank-White (NFW) dark matter profile density, which takes the form
\begin{equation}
    \rho(r(s,\ell,b)) = \rho_s\Big(\frac{r}{R_s}\Big)^{-\gamma}\Big(1+\frac{r}{R_s}\Big)^{-3+\gamma}\,,
    \label{eq:profile}
\end{equation}
\begin{equation}
    r(s,\ell,b) = \sqrt{s^2+R_\odot^2-2sR_\odot\cos \ell \cos b}\,,
\end{equation}
where $b$ and $\ell$ are the galactic coordinates.
Here, we fix the parameters $\gamma = 0.4$, $\rho_s \sim 0.57 \, {\rm GeV}/{\rm cm}^3$, $R_\odot = 8.5\, \rm kpc$, and $R_s \sim 83.24\, {\rm kpc}$ by minimizing the chi-squared distributions obtained in~\cite{Benito_2021}.\footnote{The chi-squared distribution given in~\cite{Benito_2021} when profiled over the other parameters is very flat in $R_s$ allowing values that range from 5 to 90 kpc. Another typical value that is chosen in the literature for $R_s$ is approximately 24 kpc. The resulting column densities that are relevant for this work are not significantly affected by this choice.}
Varying these parameters within their allow ranges does not significantly affect our results.
This is because, in the case of decaying dark matter, the choice of galactic profile parameters and their corresponding uncertainties does not carry as much weight as it does for the case of annihilating dark matter, where the flux is proportional to the square of the density rather than just a single power of the density.
Hence, the galactic neutrino flux can be computed as
\begin{equation}
    \frac{{\rm d}\phi^{\rm G}_\nu}{{\rm d}E_\nu {\rm d}\Omega} = \frac{1}{4\pi m_{\rm DM}\tau_{\rm DM}}\frac{{\rm d}N_{\nu}}{{\rm d}E_\nu}\int {\rm d}s \rho(s,\ell,b)\,,
    \label{eq:gal_nu}
\end{equation}
where $m_{\rm DM}$ and $\tau_{\rm DM}$ are the dark matter mass and lifetime, respectively.
The differential neutrino spectrum includes neutrino and anti-neutrinos of all the three flavours and is computed by means of the $\texttt{HDMSpectra}$ code~\cite{Bauer:2020jay}.

The extragalatic dark matter component, on the other hand, is generally taken to be isotropic.
The resulting extragalactic flux can be written as an integral over cosmological distances parametrized by redshift $z$, where we also need to account for the redshift in the neutrino energies.
This amounts to a difference between the fluxes at production and detection.
The corresponding neutrino flux at detection is:
\begin{equation}
    \frac{{\rm d}\phi^{\rm EG}_\nu}{{\rm d}E_\nu{\rm d}\Omega} = \frac{\Omega_{\rm DM}\rho_{\rm cr}}{4 \pi m_{\rm DM} \tau_{\rm DM}}\int \frac{{\rm d}z}{H(z)}\left.\frac{{\rm d}N_{\nu}}{{\rm d}E_\nu}\right|_{E'_\nu = E_\nu(1+z)} \,,
    \label{eq:egal_nu}
\end{equation}
where $H(z) = H_0\sqrt{\Omega_{\Lambda}+\Omega_m(1+z)^3}$ is the Hubble parameter and $\rho_{\rm cr}$ is the critical density of the Universe~\cite{Planck:2018vyg}.
Throughout this work we set the cosmological parameters to be: $\Omega_{\rm DM} = 0.23$, $\Omega_\Lambda = 0.77$, $\Omega_m = 0.27$, and $H_0 = 67.7~{\rm km/s/Mpc}$.
These are not the most recent values of the Planck parameters, but small changes to these should result in neither large changes to our conclusions nor changes to the final limits that we obtain.

Unlike in the case of neutrinos, the flux of gamma-rays produced from dark matter decay is heavily impacted by the intermediate interactions that arise from the subsequent propagation of photons and charged particles to Earth.
At high energies, photons interact with background radiation and produce electron-positron pairs, preventing a subset of gamma-rays from reaching the Earth.
Primary gamma-rays therefore experience an attenuation which is encoded in the optical depth, $\tau_{\gamma\gamma}$, which is a quantity that depends on the photon energy as well as the distance from the source to Earth.
Additionally, electrons and positrons undergo interactions such as Inverse Compton scattering with background photons, bremsstrahlung, and synchrotron radiation in the galactic and intergalactic magnetic field.
These processes can occur several times over the course of propagation, thus developing electromagnetic cascades that result in a secondary flux of low-energy photons at Earth~\cite{Berezinsky:1975zz,Coppi:1996ze,Venters:2010bq,Murase:2012df,Berezinsky:2016feh}.
Hence, the final photon flux is a combination of galactic and extragalactic components, each of which receives a prompt and two secondary (sec.) contribution from primary photon and electrons:
\begin{equation}
    \frac{{\rm d}\phi_{\gamma}^\alpha}{{\rm d}E_\gamma {\rm d}\Omega} = \frac{{\rm d}\phi^{\rm prompt}_{\gamma}}{{\rm d}E_\gamma {\rm d}\Omega} + \frac{{\rm d}\phi^{\rm sec.}_{\gamma \to \gamma}}{{\rm d}E_\gamma {\rm d}\Omega} + \frac{{\rm d}\phi^{\rm sec.}_{e^\pm \to \gamma}}{{\rm d}E_\gamma {\rm d}\Omega}\qquad \text{with}\quad \alpha={\rm G~or~EG}\,,
    \label{eq:gamma}
\end{equation}
where $\phi^{\rm prompt}_{\gamma}$ and $\phi^{\rm sec.}_{\gamma \to \gamma}$ are the unscattered and cascaded components of the primary gamma-ray flux $\phi^{\rm prompt}_{\gamma}$ respectively, while $\phi^{\rm sec.}_{e^\pm \to \gamma}$ is the contribution of gamma-rays from the interactions of the primary electrons.

The prompt gamma-ray flux can be similarly computed as~\cref{eq:gal_nu} and~\cref{eq:egal_nu} up to an attenuation factor also know as gamma-ray opacity, $e^{-\tau_{\gamma\gamma}}$.
As will be discussed in~\cref{sec:ebl}, the opacity is computed from the photon energy distribution in the background field and the cross sections for each of the relevant processes.
As a result, changes in our modeling of the background radiation result in differences in the observed prompt flux.
Although the spectrum of electromagnetic cascades initiated by primary photons and electrons approaches a nearly universal form consisting of two power-law components: $E_\gamma^{-1.5}$ for $E_\gamma \leq \mathcal{O}\left(0.1~{\rm GeV}\right)$, and $E_\gamma^{-1.9}$ for $E_\gamma \leq \mathcal{O}\left(10^5~{\rm GeV}\right)$~\cite{Berezinsky:2016feh},
different spectral energy distributions of the background radiation can still slightly alter the universal gamma-ray spectrum. The secondary gamma-ray flux can typically be computed by means of analytical and numerical approaches~\cite{PhysRevD.58.043004,Vladimirov:2010aq,Kachelriess:2011bi,Kalashev:2014xna,Berezinsky:2016feh,Blanco:2018bbf}.

In this analysis, we use the public simulation software \texttt{CRPropa}~\cite{AlvesBatista:2016vpy,Heiter:2017cev} to calculate the gamma-ray flux in~\cref{eq:gamma}. 
This code allows us to propagate photons and electrons over galactic and extragalactic distances with different EBL models.
In principle, such calculations require as input the differential spectrum of primary photons and electrons directly injected by dark matter decay as well as the dark matter distribution on galactic and extragalactic scales.
However, these calculations are computationally intensive, making the injection of different dark matter spectra impractical.
To overcome this difficulty we have developed a reweighting scheme that allows us to use a benchmark \texttt{CRPropa} simulation and transform it to yield the appropriate result.
Our benchmark simulation assumes a single unbroken power-law in energy that ranges from the minimum energy of interest to the largest dark matter mass considered.
To be able to efficiently reweight the galactic and extragalactic components, we produce two Monte Carlo data sets with the benchmark energy spectra which are then used to compute the galactic and extragalactic dark matter flux components.
In both cases, we inject a spectrum that is uniformly distributed over a sphere, where the radii are $\SI{100}\kpc$ and $\SI{10}\Gpc$ for the galactic and extragalactic sets, respectively.
Further details of the simulation and our Monte Carlo reweighting procedure can be found in~\Cref{simulation}.
\begin{figure}[t!]
    \centering
    \includegraphics[width = 1\textwidth]{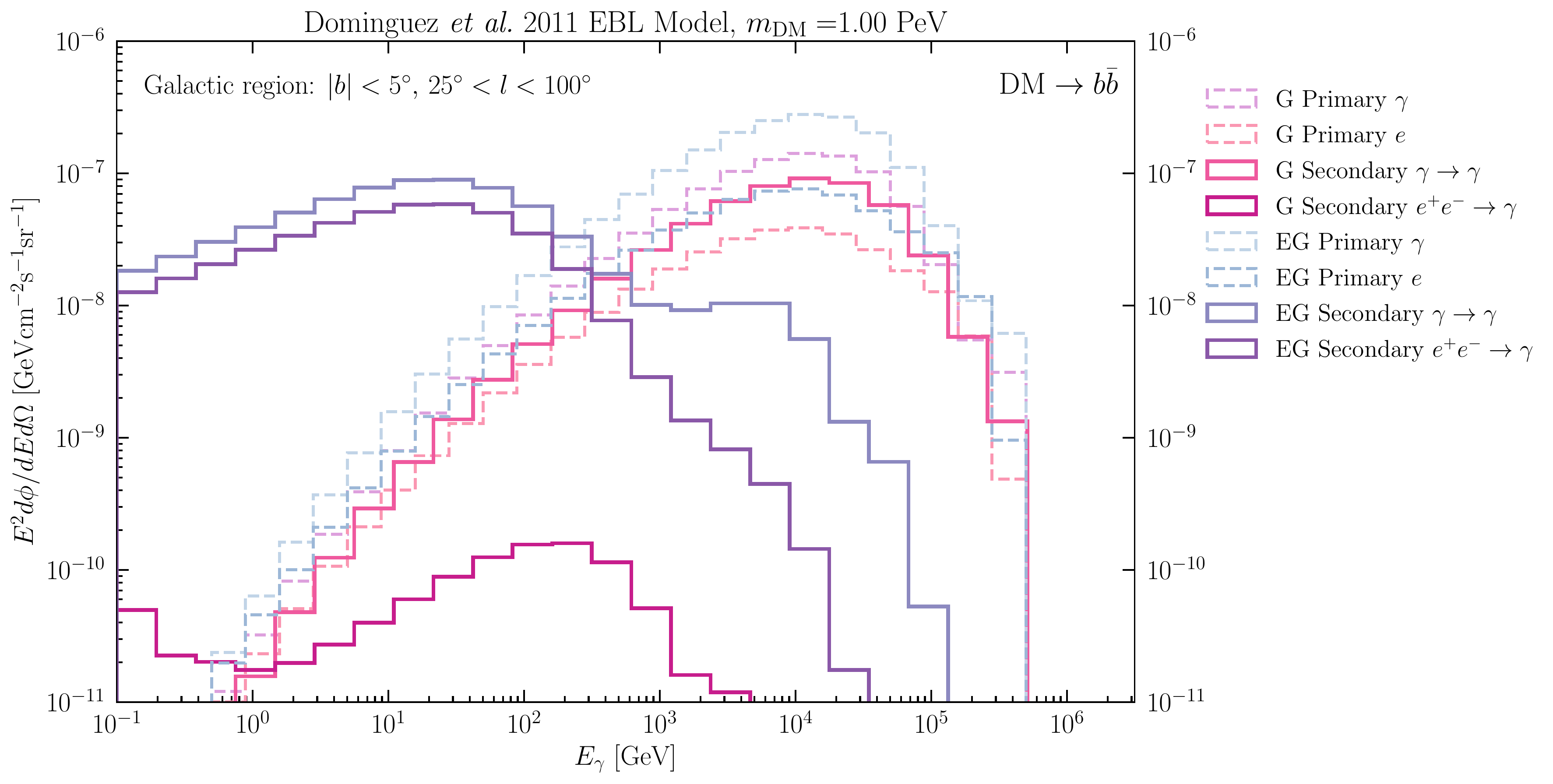}
    \caption{The different gamma-ray components from galactic (G) and extragalactic (EG) sources for $m_{\rm DM} = 1~{\rm PeV}$ and $\tau_{\rm DM} = 10^{27}~{\rm s}$.
    The dashed lines show the initial spectra for gamma-rays and $e^+e^-$ resulting from ${\rm DM}\rightarrow b\bar{b}$ decay as they appear at the source prior to propagation.
    The solid lines are the final spectra at Earth we obtain using \texttt{CRPropa} simulations and subsequently applying our reweighting procedure (see \Cref{simulation}).
    We fix the EBL spectrum according to the Dominguez \textit{et al.} model~\cite{Dominguez:2010bv}.}
    \label{fig:all_spectra}
\end{figure}

In Fig.~\ref{fig:all_spectra}, we show the different gamma-ray components described in this section, in the case of dark matter decaying to $b\bar{b}$ with a mass of 1~PeV and a lifetime of $10^{27}~{\rm s}$.
The dashed lines are the initial spectra corresponding to the gamma-rays and electron/positron final-states, directly emitted by dark matter particles.
Additionally, the solid lines represent the final gamma-ray spectra that we see at Earth after propagation.
Among these, we note that the galactic dark matter gamma-ray spectrum (red shades of color) is heavily dominated by an overall attenuation from the initial spectrum, whereas secondary production is merely a subleading effect.
This is consistent with the fact that the short propagation scales contribute to the significant suppression of gamma-ray cascade production.
In contrast, there is visible secondary production in the extragalactic components (blue shades of color), seen as the abundance at lower energies, indicating that secondary production is the dominant mechanism giving rise to gamma-rays from extragalactic dark matter decay.
Alluding to the main goal of this work, secondary production is heavily informed by the spectral energy distribution of the background radiation, suggesting that the extragalactic component will see greater effects from changes in the modeling of that background.
We elaborate on this point in the next section. 

\section{The Extragalactic Background Light\label{sec:ebl}}

The EBL, along with the Cosmic Microwave Background (CMB), is the primary target for the $\gamma - \gamma$ and $e - \gamma$ interactions.
It is produced by the radiation emitted by all the galaxies over the history of cosmic evolution. Unlike the well-measured CMB spectrum, however, there are limited ways to measure the EBL directly, and therefore it has significant uncertainties. 
In particular, there are no direct measurements of the EBL below $\sim \SI{100}\mu$m because of radiation from Zodiacal light coming from interplanetary dust radiation, which is about two orders of magnitude larger than the EBL flux~\cite{Stecker:2016fsg}. 
Collective limits obtained from direct and indirect methods show that the EBL has a two-peak structure in the spectral energy distribution from the ultraviolet and the far-infrared~\cite{Driver:2016krv,Fermi-LAT:2018lqt}.
However, the incomplete information surrounding the structure of EBL has led to the development of different models, which use different methods for estimating its structure~\cite{Dominguez:2010bv,Stecker:2016fsg,Finke:2009xi,Kneiske:2010pt,Gilmore2012,Helgason:2012xj,Inoue:2012bk,Franceschini2017,Andrews:2017ima}. 
This uncertainty propagates down to the prediction of gamma-ray fluxes from distant point-like sources~\cite{Singh:2020kmg,Saveliev:2020ynu} and the aggregated diffuse distribution of sources as is the case of dark matter.
In this analysis we consider the following benchmark models provided by Dominguez \textit{et al.} \cite{Dominguez:2010bv} and Stecker \textit{et al.} \cite{Stecker:2016fsg}. We select these as benchmark models for our analysis in part because they are motivated by observational uncertainties in the EBL and because they are supported by constraints from data.
\begin{figure}[t!]
    \centering
    \includegraphics[width = 1.0\textwidth]{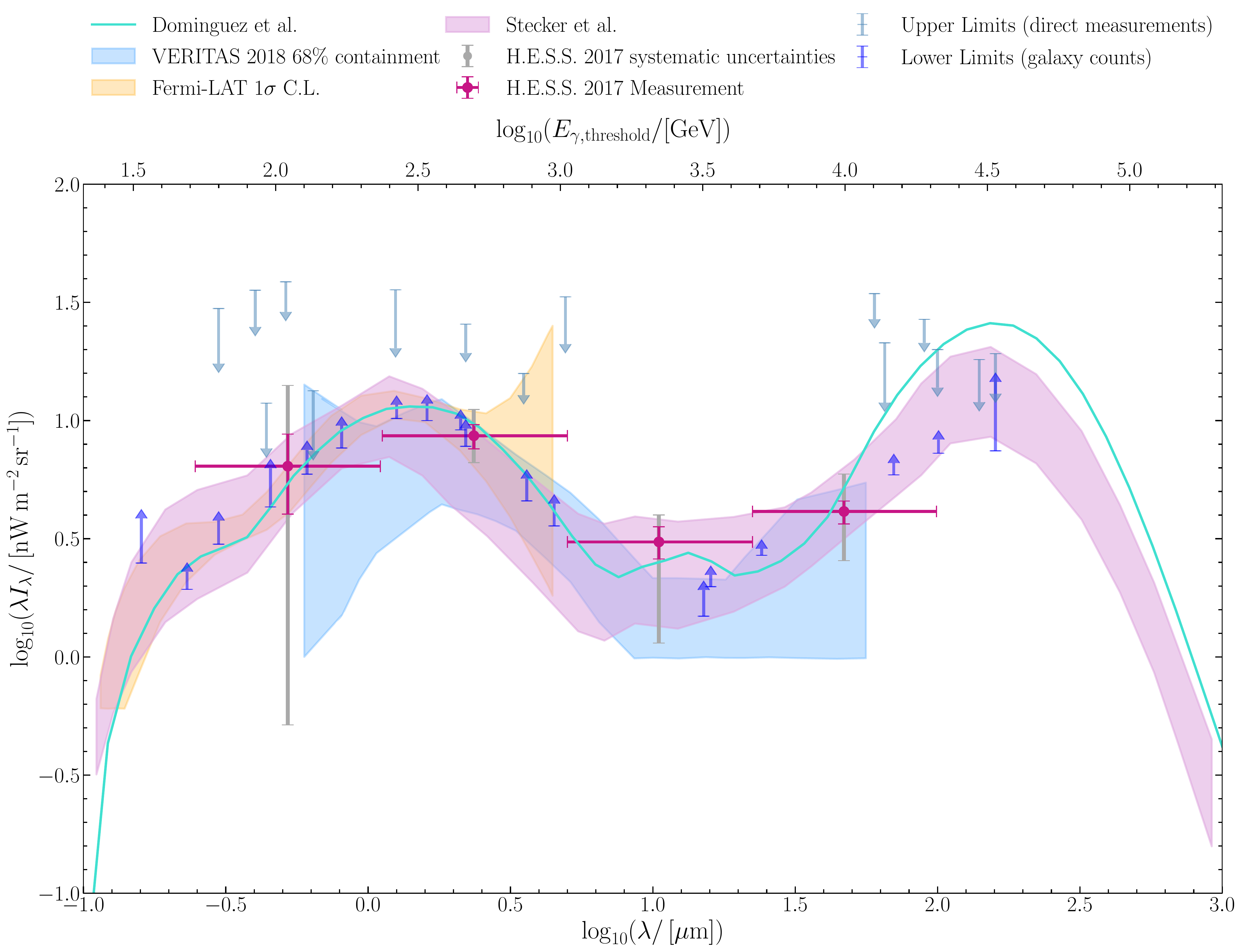}
    \caption{Spectral energy distributions (SEDs) at $z = 0$ predicted by the EBL models that we consider. Shown in purple is the 68\% confidence band obtained from Stecker \textit{et al.} Minimal and Maximal models \cite{Stecker:2016fsg}, and the cyan curve is the SED computed by Dominguez \textit{et al} \cite{Dominguez:2010bv}. The bottom axis gives the wavelength of the target photons, while the top axis recasts this in terms of the minimum (threshold) energy $E_{\gamma, \rm threshold}$ required by an incoming photon to produce an electron-positron pair. We also show several observational results for comparison, including direct constrains on the EBL in the form of upper and lower limits from direct measurements~\cite{Dwek_2013} and galaxy counts~\cite{Biteau_2015}. In addition, we show the most recent data provided by VERITAS \cite{Abeysekara_2019} and H.E.S.S~\cite{Hess2017}., and the $1\sigma$ confidence level from Fermi-LAT based on reconstructed cosmic emissivity~\cite{Fermi-LAT:2018lqt}. Note that the lower cutoff seen in the VERITAS band is artificial.}
    \label{fig:SED}
\end{figure}

The Dominguez \textit{et al.} 2011 model~\cite{Dominguez:2010bv} is an \textit{ab initio} evolution model that begins with initial conditions and evolves forward in time by using semi-analytical models of galaxy formation.
The authors adopt a novel method whereby they use the observed evolution of the rest-frame $K$-band galaxy luminosity function along with galaxy SED functions, which are based on a sample of 6000 galaxies in the range of 0.2 to 1.0 in redshift from the All-wavelength Extended Groth Strip International Survey (AEGIS).
In addition to this, the authors also use their own estimates of the SED fractions of star-forming galaxies, starburst galaxies, and active galactic nuclei (AGN) galaxies.
From this, the authors are able to extrapolate the evolution of the luminosity densities from the UV to the IR range in target wavelengths.
Comparing their results to observations, they conclude with the claim that the EBL is well-constrained in the UV to mid-IR range, but that additional constraints from IR and gamma-ray astronomy are needed in order to account for uncertainties in the far-IR region.

On the other hand, the Stecker \textit{et al.} 2016 model~\cite{Stecker:2016fsg} is a backward evolution model that begins with a present-day luminosity function and extrapolates backwards in time.
To accomplish this, they use recent deep galaxy survey data at infrared wavelengths, which is where galaxy emission arises from dust re-radiation.
In addition to this, the authors use the gamma-ray opacity from photons in the $\SI{2.7}\K$ cosmic background radiation to extend their results to energies beyond the TeV scale.
For wavelengths in the range $\SI{5}\um$ to $\SI{850}\um$, they use empirical data from the Spitzer telescope, the Herschel telescope, the Planck space telescope, and observations from the Atacama Large Millimeter Array (ALMA) and the Balloon-borne Large Aperture Submillimeter Telescope (BLAST).
Since the authors use only empirical data, their results are model independent, thereby taking into consideration the observational uncertainties of the EBL without additional assumptions about the evolution of galaxy luminosity functions, star formation rates, or galaxy evolution.
The resulting uncertainties on the EBL that they present therefore reflect the observationally determined errors that they use. 
When performing their fit to observational data, they assume that the gamma-ray opacity arises entirely from pair-production interactions with the EBL. 
Moreover, they make use of luminosity densities (LDs), which are obtained by integrating fits to observationally determined luminosity functions, to determine photon densities.
This includes fits to broken power-law functions describing galaxies as well as a double power-law fitting function.
In order to compute the 68\% limits on the LDs, the authors perform an upper and lower limit power-law function fit in the region representing the rate of decline of the LDs at redshifts beyond the peak of star formation.
In computing the upper limit, they assume a $(1+z)^{-2}$ decline in LD, while for the lower limit, they use a steeper decline of $(1+z)^{-4}$.
\begin{figure}[t!]
    \centering
    \includegraphics[width=0.45\textwidth]{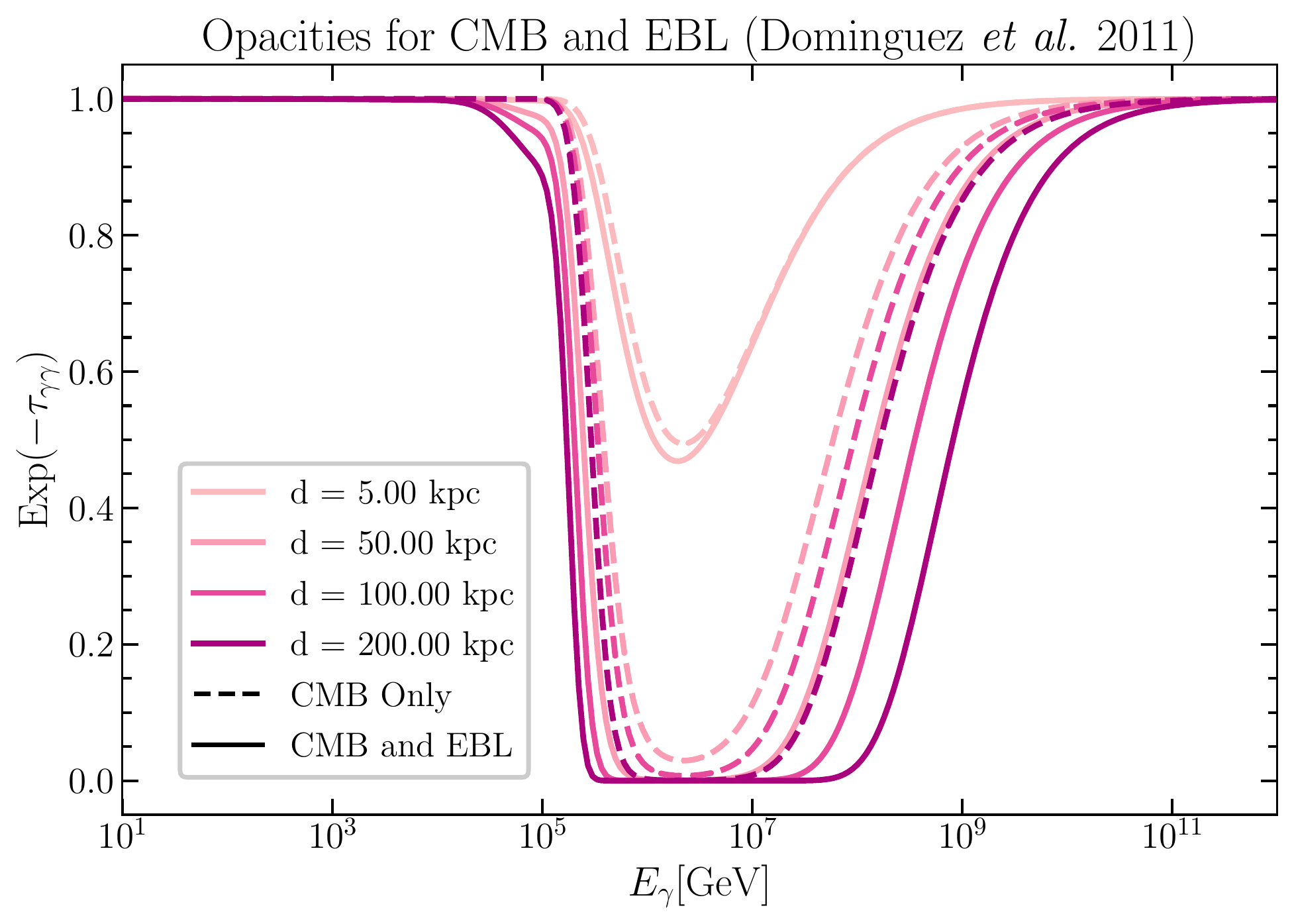}
    \includegraphics[width=0.45\textwidth]{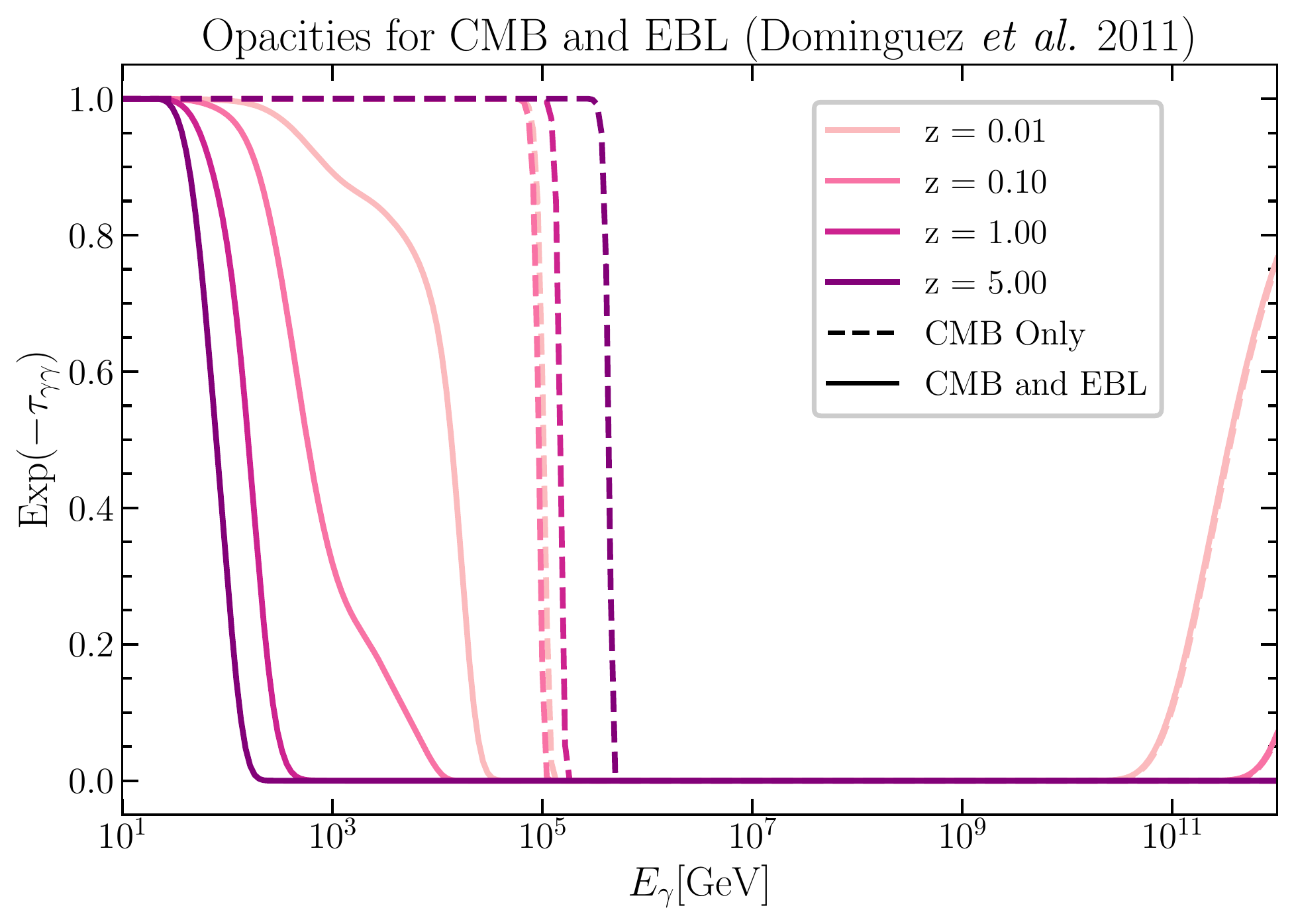} 
    \caption{Opacities calculated for sources at distances on galactic (left) and extragalactic (right) scales. The solid (dashed) lines correspond to optical depths calculated including the CMB with (without) the EBL contribution for the Dominguez \textit{et al.} model~\cite{Dominguez:2010bv}.}
    \label{fig:optdepth_1}
\end{figure}

These models predict significantly different photon number densities, encompassing the ranges allowed by uncertainties in the EBL.~\cref{fig:SED} shows the spectral energy distributions predicted by the Dominguez \textit{et al.} and Stecker \textit{et al.} models as well as data points from various observations and experiments.
These target wavelengths are in one-to-one correspondence with the minimum incoming photon energy required to produce electron-positron pairs, resulting in absorption and the development of an electromagnetic cascade.
We show this quantity, which we call $E_{\gamma, \rm threshold}$, on the top axis. In terms of the target wavelength, this is given by $E_{\gamma, \rm threshold} \sim m_e^2\lambda$, where $m_e$ is the electron mass and $\lambda$ is the target wavelength.
\begin{figure}[t!]
    \centering
    \includegraphics[width = 0.475\columnwidth]{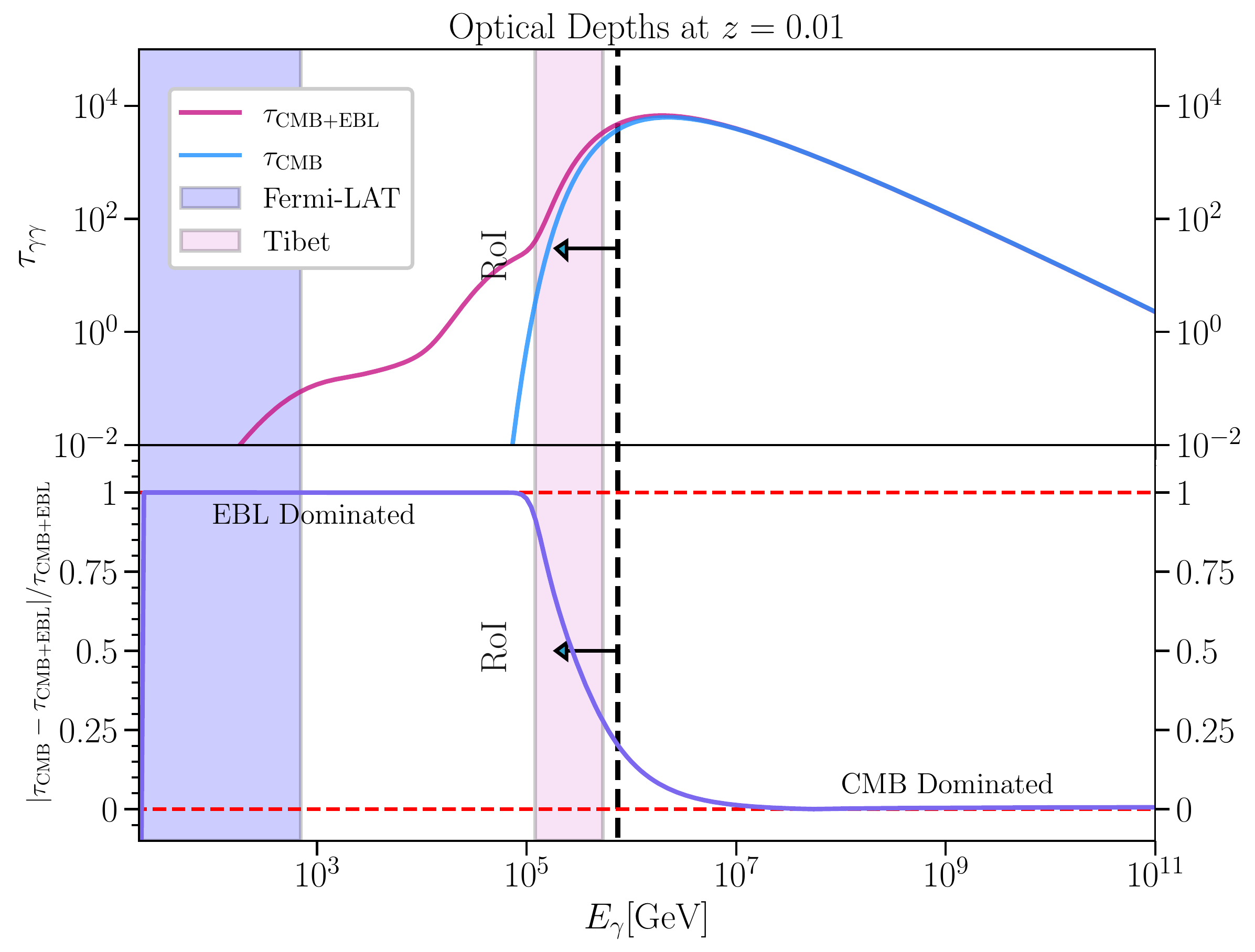}
    \hspace{0.03\columnwidth}
    \includegraphics[width = 0.465\columnwidth]{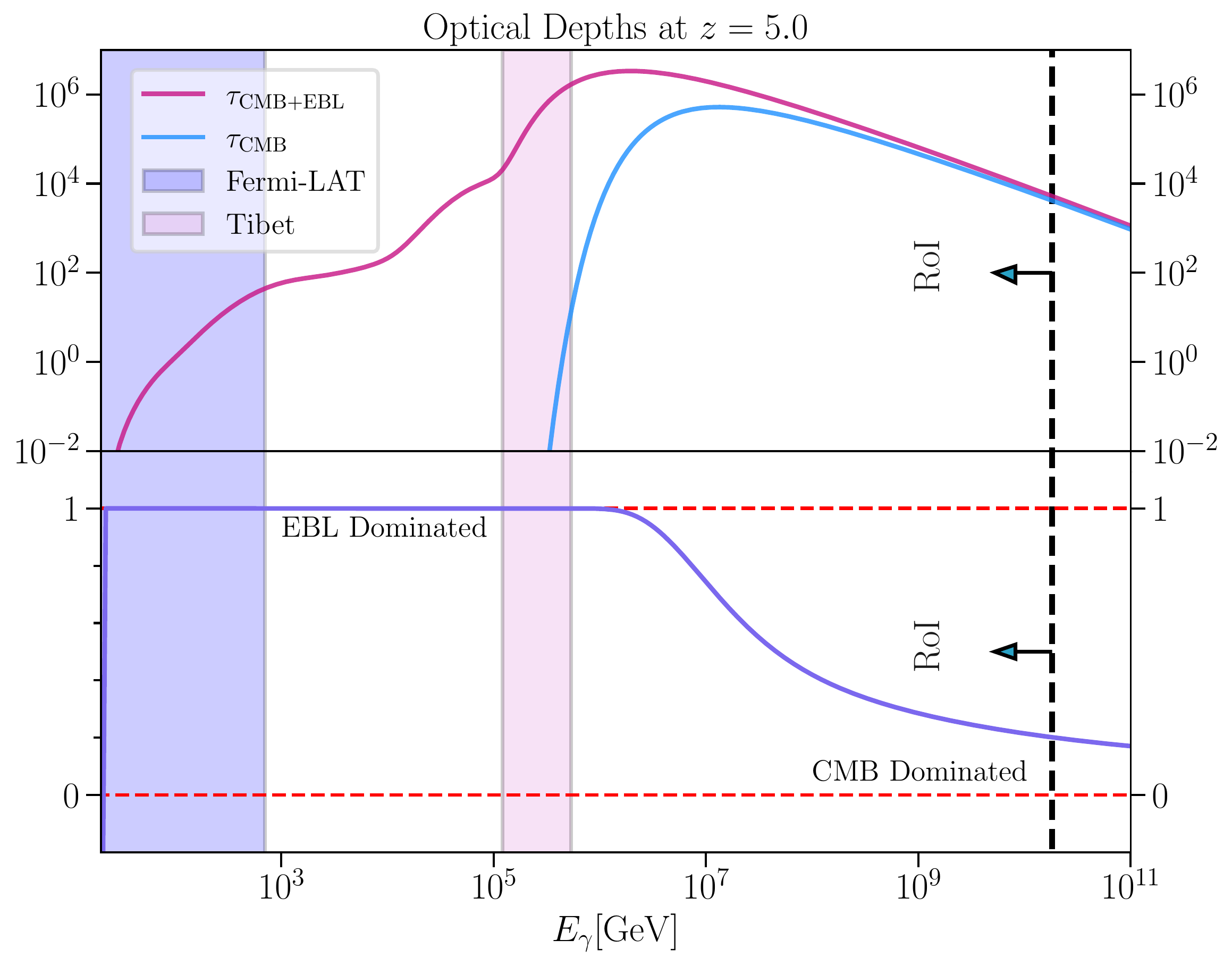}
    \caption{Optical depths (top panels) and their ratios (bottom panels) for a source of fixed distance within the galactic scale (left) and on the extragalactic scale (right).
    In the top panels, the purple (blue) lines corresponds to a target background consisting of the CMB with (without) the EBL contribution for Dominguez {\it et al.} model~\cite{Dominguez:2010bv}.
    The bottom panels shows the magnitude by which the CMB-only optical depth deviates from the CMB+EBL optical depth, illustrating the energy ranges where the EBL is significant in determining the absorption of gamma-rays as they travel to Earth.
    The dashed vertical line indicates the region that we define to be ``EBL Dominated'', or where the quantity plotted on the bottom panel is greater than 0.3.
    We call this as the ``Region of Interest'' (RoI), and it demarcates the transition between the region where the EBL is dominant according to our calculation and the region where the CMB becomes the dominant target background.}
    \label{fig:optdepth_2}
\end{figure}
\begin{figure}[t!]
    \centering
    \includegraphics[width = 1\textwidth]{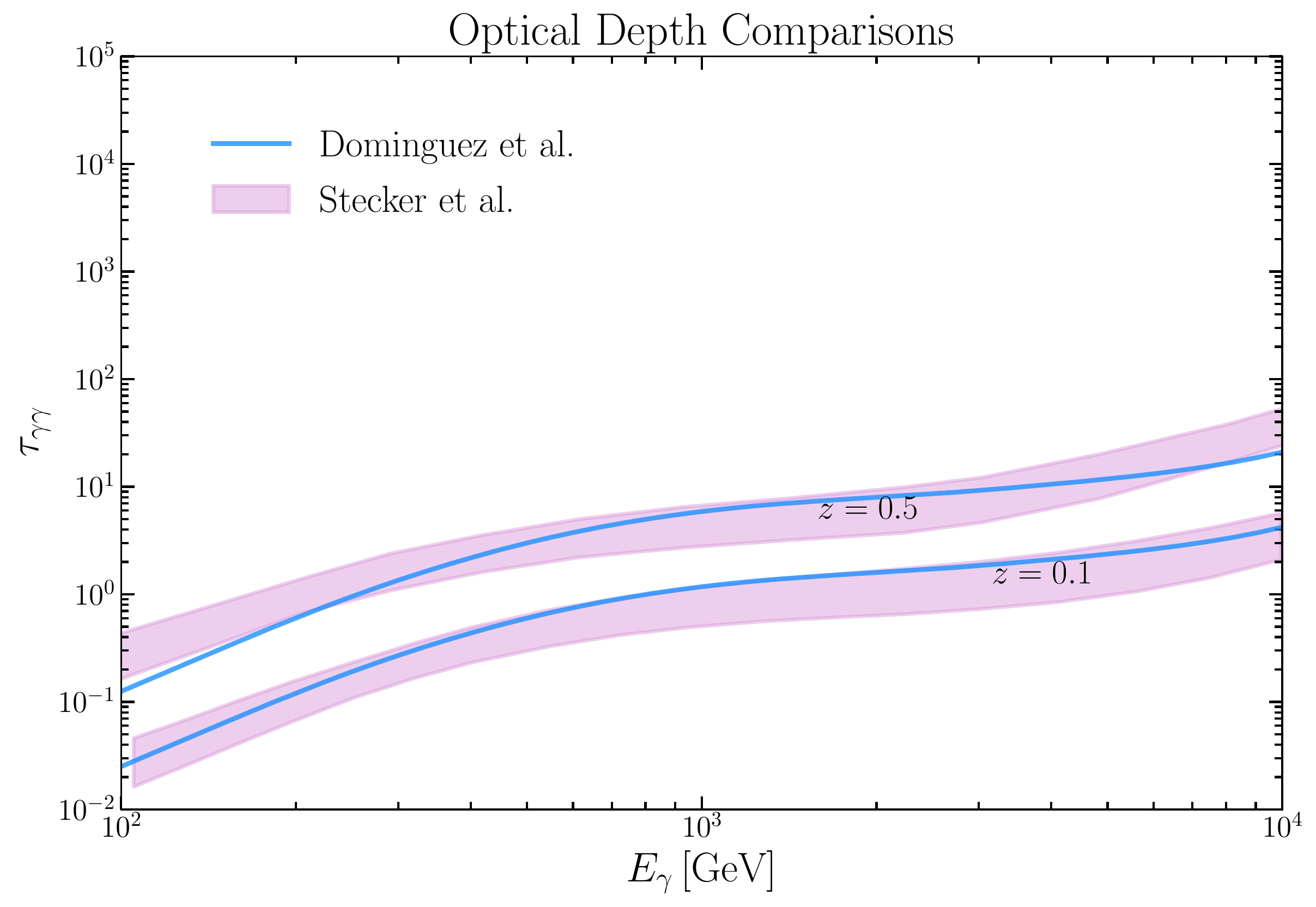}
    \caption{Optical depths at various redshifts as predicted by the Dominguez \textit{et al.} \cite{Dominguez:2010bv} and Stecker \textit{et al.} \cite{Stecker:2016fsg} EBL models. The purple band spans the range between the Stecker \textit{et al.} Minimal and Maximal models.}
    \label{fig:Tau}
\end{figure}

To understand the impact of the EBL in the gamma-ray fluxes that arrive at Earth, we can analytically calculate  the optical depth $\tau_{\gamma \gamma}$.
This quantity defines the gamma-ray opacity, $e^{-\tau_{\gamma \gamma}}$, which describes the absorption of high-energy gamma-rays due to $\gamma - \gamma$ interactions.
For photons of energy $E_{\gamma}$ emitted by a source at a redshift $z$, this is given by
\begin{equation}
    \tau_{\gamma\gamma} = \int_0^{z}{\rm d}z^{\prime}\,\frac{c}{(1+z)H(z)}\int_0^2 {\rm d}\mu\,\frac{\mu}{2}\int_0^{\infty}{\rm d}\epsilon\,\sigma_{\gamma\gamma}(E_{\gamma},\epsilon)n_\gamma(\epsilon,z^{\prime}), \quad \mu = 1-\cos\theta,
\end{equation}
where the final integral is over all possible target photon energies $\epsilon$, $\sigma_{\gamma\gamma}$ is the pair-production cross-section~\cite{Esmaili:2015xpa}, and $n_\gamma$ is the differential number density of photons in the target background, {\it i.e.} $n = n_{\rm CMB} + n_{\rm EBL}$.
When the target background is the CMB, the number density at redshift $z=0$ is given by a Planck distribution, namely
\begin{equation}
    n_{\rm CMB}(\epsilon) = \frac{\epsilon^2}{\pi^2}\frac{1}{e^{\epsilon/T_{\rm CMB}}-1},
\end{equation}
with $T_{\rm CMB} = 2.348\times 10^{-4} \, {\rm eV}$. 
In~\cref{fig:optdepth_1}, we show the gamma-ray opacity as a function of gamma-ray energy for different distances and redshifts with and without the EBL contribution according to the model by Dominguez \textit{et al.}.
At energies of around $\sim\SI{1}\PeV$, the CMB dominates the development of electromagnetic cascades, whereas the EBL starts to become relevant at lower energies.
The EBL contribution relative to the CMB, in particular, also becomes more significant with increasing distances of the emission regions of the primary gamma-rays and electrons.
As can be seen in~\cref{fig:optdepth_1} (right), which displays opacities at the extragalactic scale, the EBL contribution becomes considerably dominant at redshifts larger than $z = 0.01$; however, its effects can already be seen at the galactic scale as shown in~\cref{fig:optdepth_1} (left).

The relative contributions of the EBL can also be seen in~\cref{fig:optdepth_2}, where we show the optical depths for sources at two fixed distances. 
The bottom panels of this figure illustrate the portion of the combined optical depth ($\tau_{\rm CMB+EBL}$) that is due to interactions with the EBL, indicating the energy ranges where the EBL is relevant, denoted by an arrow as the Region of Interest (RoI). 
At both of the redshifts that we consider, the EBL becomes important in the energy ranges spanned by both Fermi-LAT and Tibet experiments, suggesting that any differences in the EBL could have implications for dark matter decay limits obtained using these datasets.
Additionally, \cref{fig:Tau} shows the optical depths that the two EBL models we consider predict for a range of redshifts and gamma-ray beam energies. Across these ranges, the opacity exhibits differences among these models.

\section{Analysis and Results\label{sec:analysis}}

The motivation behind this work is to address the previously-obtained gamma-ray dark matter constraints that disfavor the decaying dark matter hypothesis for IceCube neutrino observations and to examine the effects that the uncertainty of the EBL has on those constraints and whether this implicates previous conclusions about dark matter.
Hereafter, we obtain dark matter limits for the $b\bar{b}$ decay channel with the latest high-energy neutrino and gamma-ray data and compare our results with previous limits to illustrate the EBL effect.
We focus on the 7.5-year HESE sample, which has access to the prompt dark matter neutrino flux from the whole sky (and in particular from the Galactic center), thus providing the dominant constraints on the large-mass range.
Regarding gamma-rays, we analyze the data collected by Tibet AS$_\gamma$ and Fermi-LAT experiments.
The former has observed for the first time the Galactic emission in the sub-PeV range and is therefore sensitive primarily to the dark matter galactic component.
On the other had, the latter constrains the dark matter extragalactic component by measuring the IGRB over energies lower than a few TeV.
To derive lower limits on the dark matter lifetime $\tau_{\rm DM}$, for each of the three experiments we consider the following likelihood in the spirit of~\cite{Arguelles:2019ouk}:
\begin{equation}
   \mathcal{L}_{\rm exp}(\tau_{\rm DM},m_{\rm DM}) =  \prod_i^{n_{\rm bins}}\left\{
                \begin{array}{l r}
                  \mathcal{P}(\phi_i^{\rm exp}(E_i) | \phi_i^{\rm DM}(E_i)) & \quad(\phi_i^{\rm exp}(E_i) < \phi_i^{\rm DM}(E_i))\\
                  1 & \quad (\phi_i^{\rm exp}(E_i) \geq \phi_i^{\rm DM}(E_i)) \\
                \end{array}
              \right.,
\end{equation}
where $\phi_i^{\rm DM}(E_i;\tau_{\rm DM},m_{\rm DM})$ is the predicted energy-integrated flux on the $i$-th data bin, $\phi_i^{\rm exp}(E_i)$ the measured value by the experiment, and $\mathcal{P}$ is a Gaussian likelihood.
The width of this distribution is given by the sum of the squares of the errors, which consist of statistical errors from our Monte Carlo generation and experimental errors from the data.
We then place a constraint on $\tau_{\rm DM}$ for a given $m_{\rm DM}$ assuming Wilks' theorem with one degree of freedom.

\begin{figure}[t!]
    \centering
    \includegraphics[width = 0.75\columnwidth,height = 0.55\columnwidth]{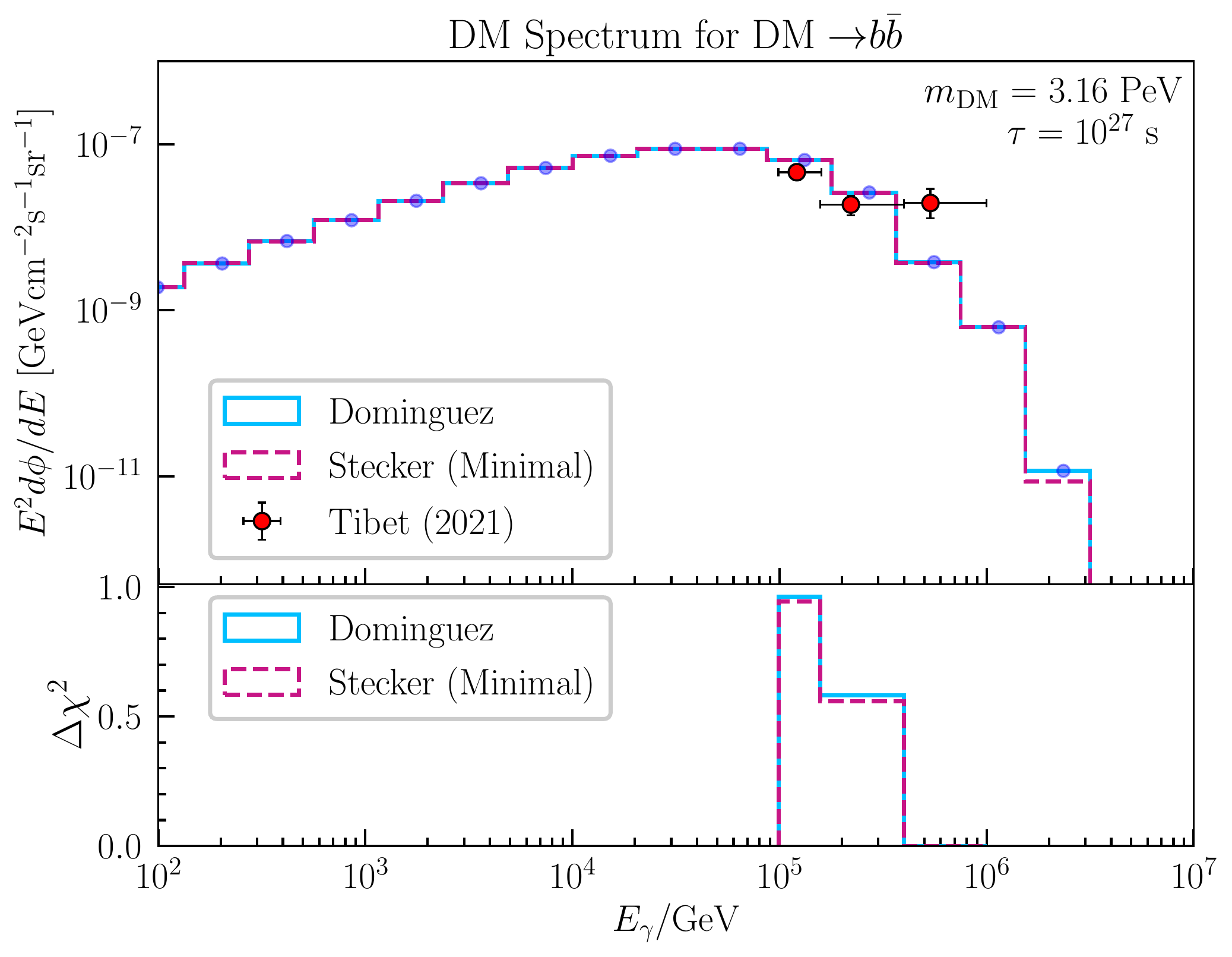}
    \caption{Final  galactic dark matter decay spectra for a chosen dark-matter mass and a lifetime of $10^{27}$ seconds. The two lines correspond to two different EBL models.
    The bottom panel shows the $\Delta \chi^2$ calculated by comparing our Monte Carlo simulations to Tibet AS$_\gamma$ data~\cite{TibetASgamma:2021tpz}.}
    \label{fig:galactic-dark-matter}
\end{figure}

The Tibet AS$_\gamma$ likelihood analysis is carried out by comparing our galactic gamma-ray dark matter distributions to the data obtained in the angular window $|b| < 5^{\circ}$ and $25^{\circ}<l<100^{\circ}$ in the three energy bins roughly defined by $[\SI{100}\TeV, \SI{160}\TeV]$, $[\SI{160}\TeV, \SI{400}\TeV]$, and $[\SI{400}\TeV, \SI{1}\PeV]$~\cite{TibetASgamma:2021tpz}.
In particular, we calculate the likelihood for five different dark matter masses ranging from 1~PeV to 10~PeV, since for smaller masses the dark matter contribution to Tibet AS$_\gamma$ data is almost negligible.
\Cref{fig:galactic-dark-matter} shows an example of the final result obtained from the combination of simulation and subsequent reweighting for $m_{\rm DM} = \SI{1}\PeV$ and $\tau_{\rm DM} = 10^{27}~{\rm s}$, using the Stecker \textit{et al.} Minimal model (dashed purple line) and the Dominguez \textit{et al.} model (solid cyan line).
As expected, the dominant contribution to the change in the final dark matter spectrum is an overall attenuation factor caused by gamma-ray absorption.
As a result of the short propagation distances, secondary gamma-rays and the inverse-Compton process are merely subleading effects, thus implying a very small uncertainty due to the choice for the EBL model.
Hence, the contributions to the $\Delta \chi^2 = -2 \ln \mathcal{L}_{\rm exp}$ (shown in the lower panel in \cref{fig:galactic-dark-matter}), and consequently the lower limits on $\tau_{\rm DM}$, are essentially the same for the different EBL models considered. This remains true across all the dark matter masses that we consider in this work.

\begin{figure}[t!]
    \centering
    \includegraphics[width = 0.75\columnwidth]{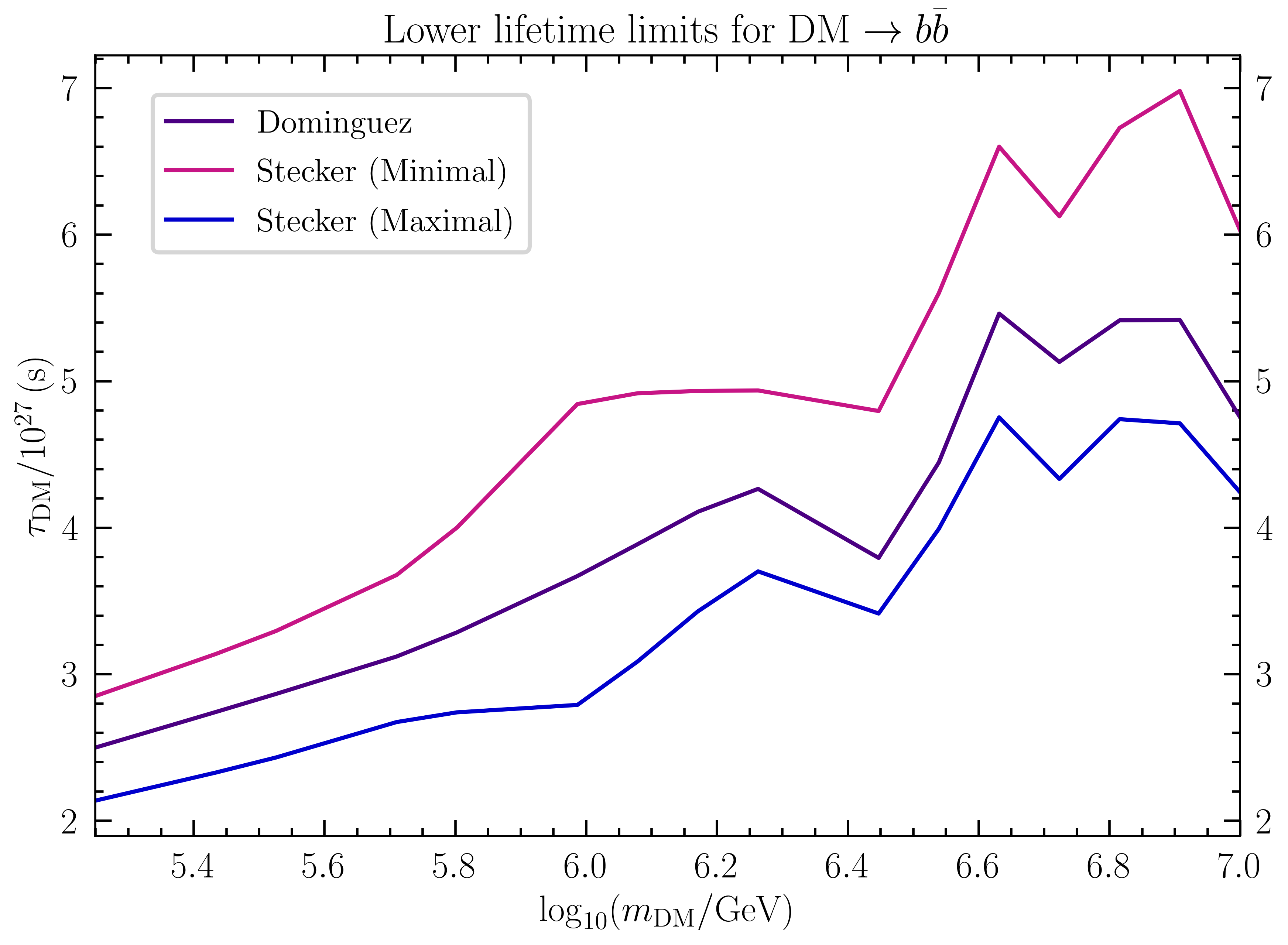}
    \caption{Fermi-LAT lower bounds at 95\% CL obtained for the three different EBL models considered in this work.}
    \label{fig:fermilat-all-bounds}
\end{figure}

For Fermi-LAT likelihood analysis, we use the P7REP IGRB HI sample, whose data are given as raw energy and positions in the sky~\cite{Fermi-LAT:2014ryh}.
Here, we ignore the spatial distribution and bin the data into 40 energy bins, equally spaced in logarithmic space from $\SI{200}\MeV$ to $\SI{2}\TeV$.
The Fermi-LAT likelihood is then calculated for eleven different dark matter masses ranging from $10^{5.25}~{\rm GeV}$ to 10~PeV.
As can be see in \cref{fig:all_spectra}, where we show an example of the final result obtained from the simulation and subsequent reweighting for a dark matter mass of $\SI{1}\PeV$ decaying to $b\bar{b}$, there is a visibly significant production of secondary gamma-rays as photons propagate to Earth, indicating that the dominant production mechanisms for visible gamma-rays from extragalactic sources are the $\gamma\rightarrow \gamma$ and $e^+e^-\rightarrow \gamma$ secondary contributions.\footnote{The comparison of these two different components to data and the contributions to the $\Delta \chi^2$ for different dark matter masses and EBL models are reported in~\Cref{app:gamma_detail}.}
This results into a significant uncertainty due to the EBL models on the dark matter constraints as shown in ~\cref{fig:fermilat-all-bounds}.
The three lines represent the lower bounds at 95\% confidence level on the dark matter lifetime for the $b\bar{b}$ decay channel obtained for different EBL models: Stecker \textit{et al.} Maximal model produces the weaker lower bound (blue line), while the Minimal one produces the strongest limit (light purple line); the model by Dominguez \textit{et al.} lies in between (dark purple line).
\begin{figure}[t!]
    \centering
    \includegraphics[width = 0.47\textwidth]{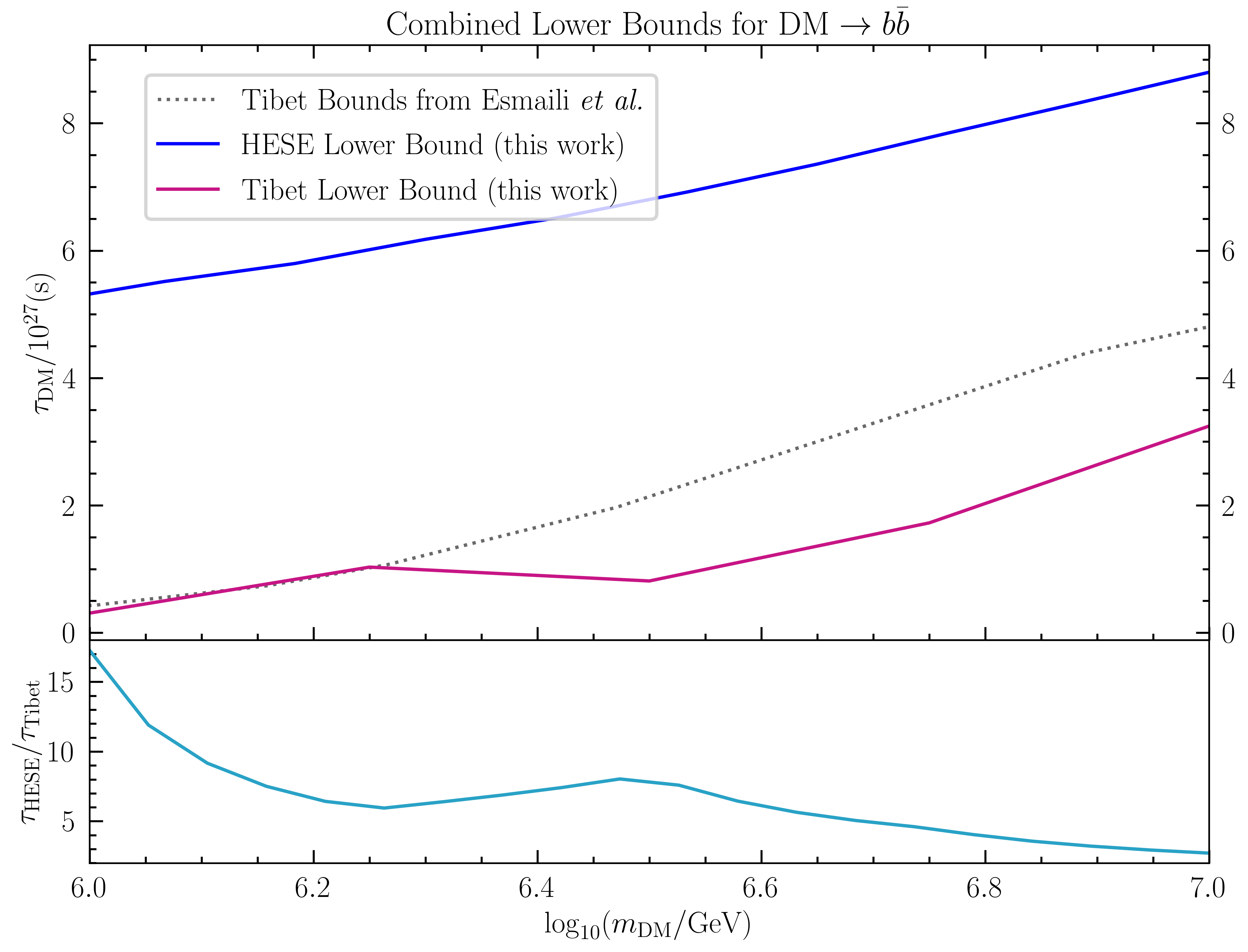}
    \includegraphics[width=0.48\textwidth]{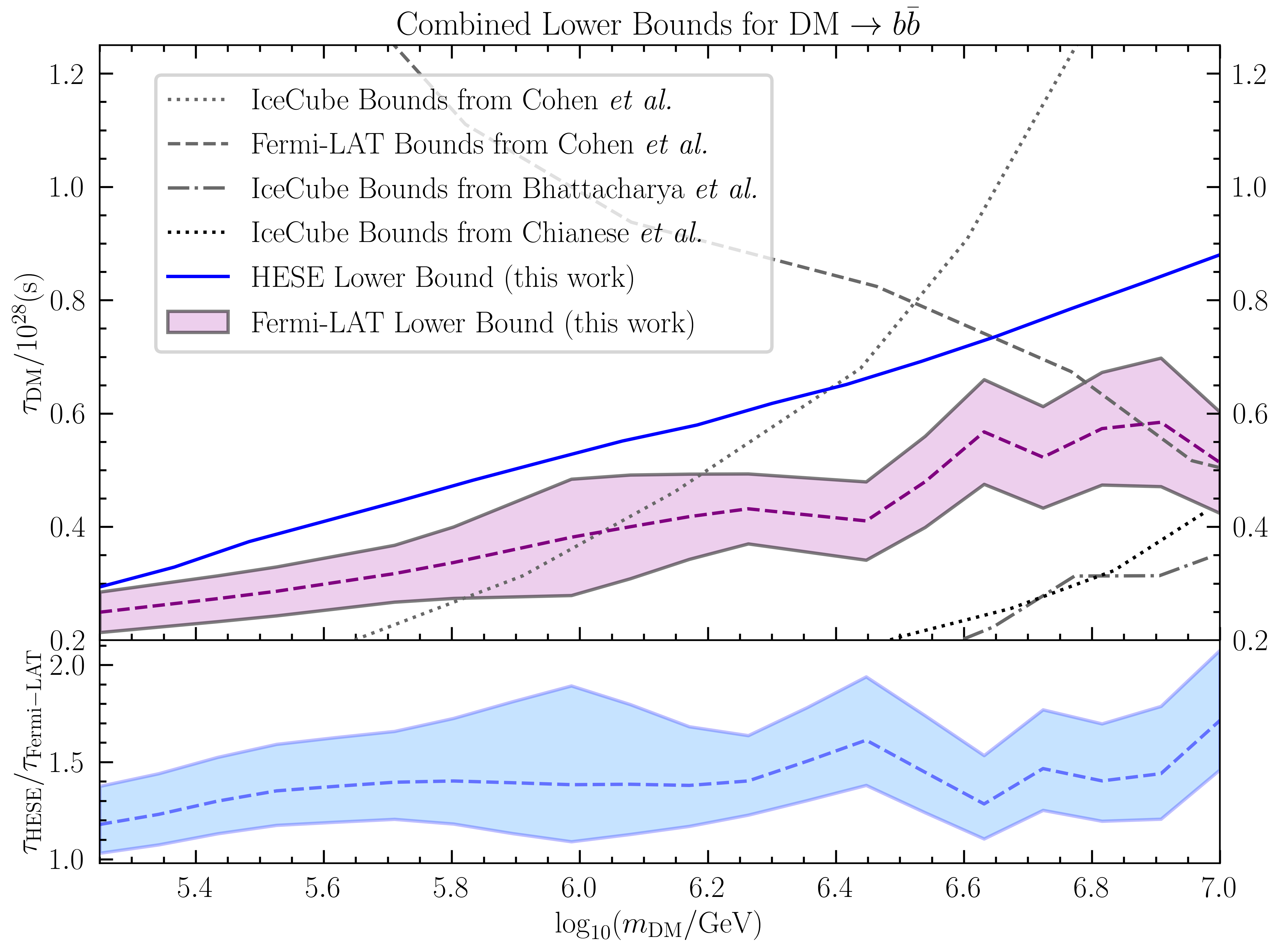}
    \caption{Lower limits at 95\% CL on the dark matter lifetime derived from IceCube 7.5-year HESE flux data (blue lines), Tibet AS$_\gamma$ data (light purple line) and Fermi-LAT diffuse measurements (dark purple band). The left and right plots refer to galactic and extragalactic gamma-ray data, respectively. The dark purple band represents the uncertainty due to different modeling of the EBL spectrum. The lower panels show the ration between the neutrino HESE limits and the galactic (left) and extragalactic (right) gamma-ray limits we obtain. Also shown with gray thin lines are previous bounds~\cite{Bhattacharya:2019ucd,Chianese:2019kyl,Cohen:2016uyg,Esmaili:2021yaw}.}
\label{fig:limits}
\end{figure}

Our results obtained using neutrinos and gamma-rays are shown in~\cref{fig:limits} for the galactic (left plot) and extragalactic (right plot) dark matter components, separately.
In particular, the dark blue lines represent the constraints placed by IceCube data, while the light (dark) purple lines in the left (right) plot corresponds to Tibet AS$_\gamma$ (Fermi-LAT) data.
In the right plot, the purple band represents the allowed range within which the EBL models we consider give predictions for the lower limit of the dark matter lifetime.
The purple dashed line within this band reflects the average taken over these EBL models.
We find that the gamma-ray limits derived with the galactic data are weaker with respect to the ones from Fermi-LAT observations (for $m_{\rm DM}$ smaller than 10~PeV), and that in general neutrino limits are stronger than gamma-ray ones.

Along with our constrains, we also plot other neutrino and gamma-ray constraints obtained outside of this work.
We emphasize that, in contrast to previously reported analyses in the literature, our analysis is background agnostic; \textit{i.e.}, we do not consider fitting to an astrophysical component for both neutrino and gamma-ray data.
However, regarding neutrino limits, we find improved constraints with respect to the works by Bhattacharya \textit{et al.}~\cite{Bhattacharya:2019ucd} and Chianese \textit{et al.}~\cite{Chianese:2019kyl}.
This is despite the fact that both of these analyses attempt to model the astrophysical background flux by means of a power-law distribution. 
Aside from this, however, here are several additional differences in our analyses, data collection, and likelihood treatments that could account for the overall improvement in our neutrino constraints.
First,~\cite{Bhattacharya:2019ucd} uses six years of runtime with energies above $\SI{10}\TeV$ and \cite{Chianese:2019kyl} uses the preliminary 7.5 years of HESE event data, whereas our analysis uses the true 7.5 year HESE flux data.
Second, for the galactic dark matter distribution we rely on a generalized NFW profile with parameters that were obtained by minimizing over the chi-squared distributions provided by the recent work~\cite{Benito_2021}.
Finally, to obtain the predicted dark matter fluxes, our analysis uses the recent code \texttt{HDMSpectra} which extends the calculations of the electroweak radiative corrections to Planck-scale dark matter masses.
For the $b\bar b$ decay channel, these processes significantly change the energy distribution of the neutrinos produced, yielding a harder neutrino spectrum compared to the ones obtained by means of other software such as \texttt{PYTHIA} 8.2~\cite{Sjostrand:2014zea} and \texttt{PPPC}~\cite{Cirelli:2010xx}(used in the analyses~\cite{Bhattacharya:2019ucd} and~\cite{Chianese:2019kyl}, respectively), which in-turn produces a stronger result.

Regarding the gamma-ray constraints, our analysis on galactic data obtains similar bounds to the ones obtained by Esmaili \textit{et al.}~\cite{Esmaili:2021yaw}. On the other hand, our results based on extragalactic data differ from the analysis performed by Cohen \textit{et al.}~\cite{Cohen:2016uyg}.
The latter considers peculiar regions around the Galactic center (see Fig.~S4 in~\cite{Cohen:2016uyg}), where the dominant contribution from decaying dark matter comes from the prompt galactic component and the secondary galactic $e^+e^-\to \gamma$ component from Inverse-Compton scattering.
We instead focus on the IGRB data from the whole sky and find that the extragalactic dark matter flux is the dominant contribution.
Moreover, Cohen \textit{et al.} model the photon background, whereas we take a model agnostic approach.
By construction, our resulting constraints are weaker than those that are background informed, as can be seen in the right plot of \cref{fig:limits}, where previous constraints are significantly stronger for masses smaller than $10^{6.5}~\si\GeV$.
However, at the largest masses, we find good agreement with~\cite{Cohen:2016uyg}, particularly when comparing to our mean prediction.\footnote{The almost flat behaviour of our limit is indeed in agreement with one placed by Cohen \textit{et al.} for the extragalactic dark matter component only (see Fig.~S13 in~\cite{Cohen:2016uyg}).}
The width of the EBL uncertainty band allows for even stronger constraints that potentially extend to higher dark matter masses.
At the highest masses that we consider, this uncertainty gives rise to deviations of up to 40\% in lifetime.
For $m_{\rm DM} = 1~{\rm PeV}$, the EBL uncertainty amounts to a $\sim 55\%$ variation for the dark matter lower bound on the lifetime.

Finally, we comment on the relative strengths of the gamma-ray and neutrino constraints obtained in this work.
These are shown in the bottom panels of~\cref{fig:limits}, where we plot the ratio between the lifetime neutrino constraints and the gamma-ray constraints.
Since the flux is inversely proportional to the lifetime, this ratio is proportional to the ratio of the dark matter fluxes between neutrinos and photons.
We find that this ratio is always greater than one in the mass range given, even when accounting for the EBL uncertainties.
This implies that in a background agnostic treatment a significant fraction of the neutrino flux observed in HESE can be accommodated with a dark matter component.
However, regarding the extragalactic gamma-ray limit, when the background is modeled, the constraint from Cohen \textit{et al.} applies and then we find that this ratio is less than one below approximately $m_{\rm DM} = 10^{6.5}~\si\GeV$, and greater than one for higher masses.
Our results highlight the importance of properly modeling the gamma-ray background in the Fermi-LAT data to derive proper constraints and, moreover, that at the highest masses, the neutrino constraints dominate over the gamma-ray limits.

\section{Conclusions \label{sec:conclusion}}

In this work, we present newly-obtained neutrino and gamma-ray limits on galactic and extragalactic dark matter decay.
We initialize dark matter decay spectra for neutrinos, gamma-rays, electrons and positrons, and we propagate these to Earth using three different models for the EBL: Dominguez \textit{et al.} and the Stecker \textit{et al.} Minimal and Maximal models.
We then compare our spectra to the 7.5-year IceCube HESE sample, the Tibet AS$_\gamma$ 2021 data, and the Fermi-LAT diffuse data.
The resulting background-agnostic analysis that we carry out yields constraints that are, taken by themselves, consistent with a decaying dark matter component explaining the HESE data, therefore providing some motivation for considering this as a partial explanation for the tension between the HESE analysis and the Northern tracks data analysis regarding the nature of the astrophysical neutrino spectrum.
\Cref{fig:limits} summarizes our results, showing that the lower bounds estimated for decaying heavy dark matter can vary by up to $\sim 55\%$ (for \SI{1}\PeV) and up to $\sim 40 \%$ at the larger mass end, due to the current uncertainty on the EBL spectrum.
In comparison to previously-obtained limits using gamma-rays and neutrinos, our constraints are in some respects better, but in other respects they can be made more robust by incorporating a background model. 

Our work motivates performing a similar investigation of the effects of the EBL on multi-messenger signals from other kinds of astrophysical sources.
Finally, modeling a gamma-ray and neutrino background is an important step in determining the extent to which high-energy neutrinos observed in samples such as HESE or the through-going muon tracks come from heavy dark matter decay.
However, this astrophysical background comes from as yet unidentified sources.

\acknowledgments

We thank Aaron C. Vincent and Sergio Palomares-Ruiz for useful comments and discussions, and Rafael Alves Batista for clarifications about \texttt{CRPropa}. CAA and BS are supported by the Faculty of Arts and Sciences of Harvard University.
Additionally, CAA thanks the Alfred P. Sloan Foundation for their support. MC is supported by the research grant number 2017W4HA7S ``NAT-NET: Neutrino and Astroparticle Theory Network'' under the program PRIN 2017 funded by the Italian Ministero dell'Universit\`a e della Ricerca (MUR) and by the research project TAsP (Theoretical Astroparticle Physics) funded by the Istituto Nazionale di Fisica Nucleare (INFN).

\bibliographystyle{jhep}
\bibliography{references}

\providecommand{\href}[2]{#2}\begingroup\raggedright\begin{thebibliography}{10}

\bibitem{Ackermann:2019ows}
M.~Ackermann et~al., \emph{{Astrophysics Uniquely Enabled by Observations of
  High-Energy Cosmic Neutrinos}}, {\emph{Bull. Am. Astron. Soc.} {\bfseries 51}
  (2019) 185} [\href{https://arxiv.org/abs/1903.04334}{{\ttfamily
  1903.04334}}].

\bibitem{Ackermann:2019cxh}
M.~Ackermann et~al., \emph{{Fundamental Physics with High-Energy Cosmic
  Neutrinos}}, {\emph{Bull. Am. Astron. Soc.} {\bfseries 51} (2019) 215}
  [\href{https://arxiv.org/abs/1903.04333}{{\ttfamily 1903.04333}}].

\bibitem{Ackermann:2022rqc}
M.~Ackermann et~al., \emph{{High-Energy and Ultra-High-Energy Neutrinos}},
  \href{https://arxiv.org/abs/2203.08096}{{\ttfamily 2203.08096}}.

\bibitem{Meszaros:2019xej}
P.~M\'esz\'aros, D.B.~Fox, C.~Hanna and K.~Murase, \emph{{Multi-Messenger
  Astrophysics}}, \href{https://doi.org/10.1038/s42254-019-0101-z}{\emph{Nature
  Rev. Phys.} {\bfseries 1} (2019) 585}
  [\href{https://arxiv.org/abs/1906.10212}{{\ttfamily 1906.10212}}].

\bibitem{AlvesBatista:2021gzc}
R.~Alves~Batista et~al., \emph{{EuCAPT White Paper: Opportunities and
  Challenges for Theoretical Astroparticle Physics in the Next Decade}},
  \href{https://arxiv.org/abs/2110.10074}{{\ttfamily 2110.10074}}.

\bibitem{IceCube:2020wum}
{\scshape IceCube} collaboration, \emph{{The IceCube high-energy starting event
  sample: Description and flux characterization with 7.5 years of data}},
  \href{https://doi.org/10.1103/PhysRevD.104.022002}{\emph{Phys. Rev. D}
  {\bfseries 104} (2021) 022002}
  [\href{https://arxiv.org/abs/2011.03545}{{\ttfamily 2011.03545}}].

\bibitem{Loeb:2006tw}
A.~Loeb and E.~Waxman, \emph{{The Cumulative background of high energy
  neutrinos from starburst galaxies}},
  \href{https://doi.org/10.1088/1475-7516/2006/05/003}{\emph{JCAP} {\bfseries
  05} (2006) 003} [\href{https://arxiv.org/abs/astro-ph/0601695}{{\ttfamily
  astro-ph/0601695}}].

\bibitem{Ambrosone:2020evo}
A.~Ambrosone, M.~Chianese, D.F.G.~Fiorillo, A.~Marinelli, G.~Miele and
  O.~Pisanti, \emph{{Starburst galaxies strike back: a multi-messenger analysis
  with Fermi-LAT and IceCube data}},
  \href{https://doi.org/10.1093/mnras/stab659}{\emph{Mon. Not. Roy. Astron.
  Soc.} {\bfseries 503} (2021) 4032}
  [\href{https://arxiv.org/abs/2011.02483}{{\ttfamily 2011.02483}}].

\bibitem{Winter:2013cla}
W.~Winter, \emph{{Photohadronic Origin of the TeV-PeV Neutrinos Observed in
  IceCube}}, \href{https://doi.org/10.1103/PhysRevD.88.083007}{\emph{Phys. Rev.
  D} {\bfseries 88} (2013) 083007}
  [\href{https://arxiv.org/abs/1307.2793}{{\ttfamily 1307.2793}}].

\bibitem{Fiorillo:2021hty}
D.F.G.~Fiorillo, A.~Van~Vliet, S.~Morisi and W.~Winter, \emph{{Unified thermal
  model for photohadronic neutrino production in astrophysical sources}},
  \href{https://doi.org/10.1088/1475-7516/2021/07/028}{\emph{JCAP} {\bfseries
  07} (2021) 028} [\href{https://arxiv.org/abs/2103.16577}{{\ttfamily
  2103.16577}}].

\bibitem{MAGIC:2018sak}
{\scshape MAGIC} collaboration, \emph{{The blazar TXS 0506+056 associated with
  a high-energy neutrino: insights into extragalactic jets and cosmic ray
  acceleration}},
  \href{https://doi.org/10.3847/2041-8213/aad083}{\emph{Astrophys. J. Lett.}
  {\bfseries 863} (2018) L10}
  [\href{https://arxiv.org/abs/1807.04300}{{\ttfamily 1807.04300}}].

\bibitem{IceCube:2018dnn}
{\scshape IceCube, Fermi-LAT, MAGIC, AGILE, ASAS-SN, HAWC, H.E.S.S., INTEGRAL,
  Kanata, Kiso, Kapteyn, Liverpool Telescope, Subaru, Swift NuSTAR, VERITAS,
  VLA/17B-403} collaboration, \emph{{Multimessenger observations of a flaring
  blazar coincident with high-energy neutrino IceCube-170922A}},
  \href{https://doi.org/10.1126/science.aat1378}{\emph{Science} {\bfseries 361}
  (2018) eaat1378} [\href{https://arxiv.org/abs/1807.08816}{{\ttfamily
  1807.08816}}].

\bibitem{IceCube:2018cha}
{\scshape IceCube} collaboration, \emph{{Neutrino emission from the direction
  of the blazar TXS 0506+056 prior to the IceCube-170922A alert}},
  \href{https://doi.org/10.1126/science.aat2890}{\emph{Science} {\bfseries 361}
  (2018) 147} [\href{https://arxiv.org/abs/1807.08794}{{\ttfamily
  1807.08794}}].

\bibitem{IceCube:2021uhz}
{\scshape IceCube} collaboration, \emph{{Improved Characterization of the
  Astrophysical Muon-Neutrino Flux with 9.5 Years of IceCube Data}},
  \href{https://arxiv.org/abs/2111.10299}{{\ttfamily 2111.10299}}.

\bibitem{Waxman_1998}
E.~Waxman and J.~Bahcall, \emph{High energy neutrinos from astrophysical
  sources: An upper bound},
  \href{https://doi.org/10.1103/physrevd.59.023002}{\emph{Physical Review D}
  {\bfseries 59} (1998) }.

\bibitem{Chen:2014gxa}
C.-Y.~Chen, P.S.~Bhupal~Dev and A.~Soni, \emph{{Two-component flux explanation
  for the high energy neutrino events at IceCube}},
  \href{https://doi.org/10.1103/PhysRevD.92.073001}{\emph{Phys. Rev. D}
  {\bfseries 92} (2015) 073001}
  [\href{https://arxiv.org/abs/1411.5658}{{\ttfamily 1411.5658}}].

\bibitem{IceCube:2015gsk}
{\scshape IceCube} collaboration, \emph{{A combined maximum-likelihood analysis
  of the high-energy astrophysical neutrino flux measured with IceCube}},
  \href{https://doi.org/10.1088/0004-637X/809/1/98}{\emph{Astrophys. J.}
  {\bfseries 809} (2015) 98}
  [\href{https://arxiv.org/abs/1507.03991}{{\ttfamily 1507.03991}}].

\bibitem{Gaggero:2015xza}
D.~Gaggero, D.~Grasso, A.~Marinelli, A.~Urbano and M.~Valli, \emph{{The
  gamma-ray and neutrino sky: A consistent picture of Fermi-LAT, Milagro, and
  IceCube results}},
  \href{https://doi.org/10.1088/2041-8205/815/2/L25}{\emph{Astrophys. J. Lett.}
  {\bfseries 815} (2015) L25}
  [\href{https://arxiv.org/abs/1504.00227}{{\ttfamily 1504.00227}}].

\bibitem{Chianese:2016opp}
M.~Chianese, G.~Miele, S.~Morisi and E.~Vitagliano, \emph{{Low energy IceCube
  data and a possible Dark Matter related excess}},
  \href{https://doi.org/10.1016/j.physletb.2016.03.084}{\emph{Phys. Lett. B}
  {\bfseries 757} (2016) 251}
  [\href{https://arxiv.org/abs/1601.02934}{{\ttfamily 1601.02934}}].

\bibitem{Palladino:2016zoe}
A.~Palladino and F.~Vissani, \emph{{Extragalactic plus Galactic model for
  IceCube neutrino events}},
  \href{https://doi.org/10.3847/0004-637X/826/2/185}{\emph{Astrophys. J.}
  {\bfseries 826} (2016) 185}
  [\href{https://arxiv.org/abs/1601.06678}{{\ttfamily 1601.06678}}].

\bibitem{Vincent:2016nut}
A.C.~Vincent, S.~Palomares-Ruiz and O.~Mena, \emph{{Analysis of the 4-year
  IceCube high-energy starting events}},
  \href{https://doi.org/10.1103/PhysRevD.94.023009}{\emph{Phys. Rev. D}
  {\bfseries 94} (2016) 023009}
  [\href{https://arxiv.org/abs/1605.01556}{{\ttfamily 1605.01556}}].

\bibitem{Palladino:2016xsy}
A.~Palladino, M.~Spurio and F.~Vissani, \emph{{On the IceCube spectral
  anomaly}}, \href{https://doi.org/10.1088/1475-7516/2016/12/045}{\emph{JCAP}
  {\bfseries 12} (2016) 045}
  [\href{https://arxiv.org/abs/1610.07015}{{\ttfamily 1610.07015}}].

\bibitem{Anchordoqui:2016ewn}
L.A.~Anchordoqui, M.M.~Block, L.~Durand, P.~Ha, J.F.~Soriano and T.J.~Weiler,
  \emph{{Evidence for a break in the spectrum of astrophysical neutrinos}},
  \href{https://doi.org/10.1103/PhysRevD.95.083009}{\emph{Phys. Rev. D}
  {\bfseries 95} (2017) 083009}
  [\href{https://arxiv.org/abs/1611.07905}{{\ttfamily 1611.07905}}].

\bibitem{Denton:2017csz}
P.B.~Denton, D.~Marfatia and T.J.~Weiler, \emph{{The Galactic Contribution to
  IceCube's Astrophysical Neutrino Flux}},
  \href{https://doi.org/10.1088/1475-7516/2017/08/033}{\emph{JCAP} {\bfseries
  08} (2017) 033} [\href{https://arxiv.org/abs/1703.09721}{{\ttfamily
  1703.09721}}].

\bibitem{Chianese:2017jfa}
M.~Chianese, R.~Mele, G.~Miele, P.~Migliozzi and S.~Morisi, \emph{{Use of
  ANTARES and IceCube Data to Constrain a Single Power-law Neutrino Flux}},
  \href{https://doi.org/10.3847/1538-4357/aa97e6}{\emph{Astrophys. J.}
  {\bfseries 851} (2017) 36}
  [\href{https://arxiv.org/abs/1707.05168}{{\ttfamily 1707.05168}}].

\bibitem{Palladino:2018evm}
A.~Palladino and W.~Winter, \emph{{A multi-component model for observed
  astrophysical neutrinos}},
  \href{https://doi.org/10.3204/PUBDB-2018-01376}{\emph{Astron. Astrophys.}
  {\bfseries 615} (2018) A168}
  [\href{https://arxiv.org/abs/1801.07277}{{\ttfamily 1801.07277}}].

\bibitem{Sui:2018bbh}
Y.~Sui and P.S.~Bhupal~Dev, \emph{{A Combined Astrophysical and Dark Matter
  Interpretation of the IceCube HESE and Throughgoing Muon Events}},
  \href{https://doi.org/10.1088/1475-7516/2018/07/020}{\emph{JCAP} {\bfseries
  07} (2018) 020} [\href{https://arxiv.org/abs/1804.04919}{{\ttfamily
  1804.04919}}].

\bibitem{Feldstein:2013kka}
B.~Feldstein, A.~Kusenko, S.~Matsumoto and T.T.~Yanagida, \emph{{Neutrinos at
  IceCube from Heavy Decaying Dark Matter}},
  \href{https://doi.org/10.1103/PhysRevD.88.015004}{\emph{Phys. Rev. D}
  {\bfseries 88} (2013) 015004}
  [\href{https://arxiv.org/abs/1303.7320}{{\ttfamily 1303.7320}}].

\bibitem{Esmaili:2013gha}
A.~Esmaili and P.D.~Serpico, \emph{{Are IceCube neutrinos unveiling PeV-scale
  decaying dark matter?}},
  \href{https://doi.org/10.1088/1475-7516/2013/11/054}{\emph{JCAP} {\bfseries
  11} (2013) 054} [\href{https://arxiv.org/abs/1308.1105}{{\ttfamily
  1308.1105}}].

\bibitem{Bai:2013nga}
Y.~Bai, R.~Lu and J.~Salvado, \emph{{Geometric Compatibility of IceCube TeV-PeV
  Neutrino Excess and its Galactic Dark Matter Origin}},
  \href{https://doi.org/10.1007/JHEP01(2016)161}{\emph{JHEP} {\bfseries 01}
  (2016) 161} [\href{https://arxiv.org/abs/1311.5864}{{\ttfamily 1311.5864}}].

\bibitem{Bhattacharya:2014vwa}
A.~Bhattacharya, M.H.~Reno and I.~Sarcevic, \emph{{Reconciling neutrino flux
  from heavy dark matter decay and recent events at IceCube}},
  \href{https://doi.org/10.1007/JHEP06(2014)110}{\emph{JHEP} {\bfseries 06}
  (2014) 110} [\href{https://arxiv.org/abs/1403.1862}{{\ttfamily 1403.1862}}].

\bibitem{Esmaili:2014rma}
A.~Esmaili, S.K.~Kang and P.D.~Serpico, \emph{{IceCube events and decaying dark
  matter: hints and constraints}},
  \href{https://doi.org/10.1088/1475-7516/2014/12/054}{\emph{JCAP} {\bfseries
  12} (2014) 054} [\href{https://arxiv.org/abs/1410.5979}{{\ttfamily
  1410.5979}}].

\bibitem{Kachelriess:2018rty}
M.~Kachelriess, O.E.~Kalashev and M.Y.~Kuznetsov, \emph{{Heavy decaying dark
  matter and IceCube high energy neutrinos}},
  \href{https://doi.org/10.1103/PhysRevD.98.083016}{\emph{Phys. Rev. D}
  {\bfseries 98} (2018) 083016}
  [\href{https://arxiv.org/abs/1805.04500}{{\ttfamily 1805.04500}}].

\bibitem{Denton:2018aml}
P.B.~Denton and I.~Tamborra, \emph{{Invisible Neutrino Decay Could Resolve
  IceCube\textquoteright{}s Track and Cascade Tension}},
  \href{https://doi.org/10.1103/PhysRevLett.121.121802}{\emph{Phys. Rev. Lett.}
  {\bfseries 121} (2018) 121802}
  [\href{https://arxiv.org/abs/1805.05950}{{\ttfamily 1805.05950}}].

\bibitem{Bhattacharya:2019ucd}
A.~Bhattacharya, A.~Esmaili, S.~Palomares-Ruiz and I.~Sarcevic, \emph{{Update
  on decaying and annihilating heavy dark matter with the 6-year IceCube HESE
  data}}, \href{https://doi.org/10.1088/1475-7516/2019/05/051}{\emph{JCAP}
  {\bfseries 05} (2019) 051}
  [\href{https://arxiv.org/abs/1903.12623}{{\ttfamily 1903.12623}}].

\bibitem{Chianese:2019kyl}
M.~Chianese, D.F.G.~Fiorillo, G.~Miele, S.~Morisi and O.~Pisanti,
  \emph{{Decaying dark matter at IceCube and its signature on High Energy gamma
  experiments}},
  \href{https://doi.org/10.1088/1475-7516/2019/11/046}{\emph{JCAP} {\bfseries
  11} (2019) 046} [\href{https://arxiv.org/abs/1907.11222}{{\ttfamily
  1907.11222}}].

\bibitem{Dekker:2019gpe}
A.~Dekker, M.~Chianese and S.~Ando, \emph{{Probing dark matter signals in
  neutrino telescopes through angular power spectrum}},
  \href{https://doi.org/10.1088/1475-7516/2020/09/007}{\emph{JCAP} {\bfseries
  09} (2020) 007} [\href{https://arxiv.org/abs/1910.12917}{{\ttfamily
  1910.12917}}].

\bibitem{IceCube:2018tkk}
{\scshape IceCube} collaboration, \emph{{Search for neutrinos from decaying
  dark matter with IceCube}},
  \href{https://doi.org/10.1140/epjc/s10052-018-6273-3}{\emph{Eur. Phys. J. C}
  {\bfseries 78} (2018) 831}
  [\href{https://arxiv.org/abs/1804.03848}{{\ttfamily 1804.03848}}].

\bibitem{Arguelles:2019boy}
{\scshape IceCube} collaboration, \emph{{Searches for Connections Between Dark
  Matter and Neutrinos with the IceCube High-Energy Starting Event Sample}},
  \href{https://doi.org/10.22323/1.358.0839}{\emph{PoS} {\bfseries ICRC2019}
  (2020) 839} [\href{https://arxiv.org/abs/1907.11193}{{\ttfamily
  1907.11193}}].

\bibitem{IceCube:2021sog}
{\scshape IceCube} collaboration, \emph{{A Search for Neutrinos from Decaying
  Dark Matter in Galaxy Clusters and Galaxies with IceCube}},
  \href{https://doi.org/10.22323/1.395.0506}{\emph{PoS} {\bfseries ICRC2021}
  (2021) 506} [\href{https://arxiv.org/abs/2107.11527}{{\ttfamily
  2107.11527}}].

\bibitem{Ciafaloni:2010ti}
P.~Ciafaloni, D.~Comelli, A.~Riotto, F.~Sala, A.~Strumia and A.~Urbano,
  \emph{{Weak Corrections are Relevant for Dark Matter Indirect Detection}},
  \href{https://doi.org/10.1088/1475-7516/2011/03/019}{\emph{JCAP} {\bfseries
  03} (2011) 019} [\href{https://arxiv.org/abs/1009.0224}{{\ttfamily
  1009.0224}}].

\bibitem{Buch:2015iya}
J.~Buch, M.~Cirelli, G.~Giesen and M.~Taoso, \emph{{PPPC 4 DM secondary: A Poor
  Particle Physicist Cookbook for secondary radiation from Dark Matter}},
  \href{https://doi.org/10.1088/1475-7516/2015/9/037}{\emph{JCAP} {\bfseries
  09} (2015) 037} [\href{https://arxiv.org/abs/1505.01049}{{\ttfamily
  1505.01049}}].

\bibitem{Bauer:2020jay}
C.W.~Bauer, N.L.~Rodd and B.R.~Webber, \emph{{Dark matter spectra from the
  electroweak to the Planck scale}},
  \href{https://doi.org/10.1007/JHEP06(2021)121}{\emph{JHEP} {\bfseries 06}
  (2021) 121} [\href{https://arxiv.org/abs/2007.15001}{{\ttfamily
  2007.15001}}].

\bibitem{Murase:2012xs}
K.~Murase and J.F.~Beacom, \emph{{Constraining Very Heavy Dark Matter Using
  Diffuse Backgrounds of Neutrinos and Cascaded Gamma Rays}},
  \href{https://doi.org/10.1088/1475-7516/2012/10/043}{\emph{JCAP} {\bfseries
  10} (2012) 043} [\href{https://arxiv.org/abs/1206.2595}{{\ttfamily
  1206.2595}}].

\bibitem{Murase:2015gea}
K.~Murase, R.~Laha, S.~Ando and M.~Ahlers, \emph{{Testing the Dark Matter
  Scenario for PeV Neutrinos Observed in IceCube}},
  \href{https://doi.org/10.1103/PhysRevLett.115.071301}{\emph{Phys. Rev. Lett.}
  {\bfseries 115} (2015) 071301}
  [\href{https://arxiv.org/abs/1503.04663}{{\ttfamily 1503.04663}}].

\bibitem{Esmaili:2015xpa}
A.~Esmaili and P.D.~Serpico, \emph{{Gamma-ray bounds from EAS detectors and
  heavy decaying dark matter constraints}},
  \href{https://doi.org/10.1088/1475-7516/2015/10/014}{\emph{JCAP} {\bfseries
  10} (2015) 014} [\href{https://arxiv.org/abs/1505.06486}{{\ttfamily
  1505.06486}}].

\bibitem{Kalashev:2016cre}
O.K.~Kalashev and M.Y.~Kuznetsov, \emph{{Constraining heavy decaying dark
  matter with the high energy gamma-ray limits}},
  \href{https://doi.org/10.1103/PhysRevD.94.063535}{\emph{Phys. Rev. D}
  {\bfseries 94} (2016) 063535}
  [\href{https://arxiv.org/abs/1606.07354}{{\ttfamily 1606.07354}}].

\bibitem{Cohen:2016uyg}
T.~Cohen, K.~Murase, N.L.~Rodd, B.R.~Safdi and Y.~Soreq,
  \emph{{\ensuremath{\gamma} -ray Constraints on Decaying Dark Matter and
  Implications for IceCube}},
  \href{https://doi.org/10.1103/PhysRevLett.119.021102}{\emph{Phys. Rev. Lett.}
  {\bfseries 119} (2017) 021102}
  [\href{https://arxiv.org/abs/1612.05638}{{\ttfamily 1612.05638}}].

\bibitem{HAWC:2017udy}
{\scshape HAWC} collaboration, \emph{{A Search for Dark Matter in the Galactic
  Halo with HAWC}},
  \href{https://doi.org/10.1088/1475-7516/2018/02/049}{\emph{JCAP} {\bfseries
  02} (2018) 049} [\href{https://arxiv.org/abs/1710.10288}{{\ttfamily
  1710.10288}}].

\bibitem{Blanco:2018esa}
C.~Blanco and D.~Hooper, \emph{{Constraints on Decaying Dark Matter from the
  Isotropic Gamma-Ray Background}},
  \href{https://doi.org/10.1088/1475-7516/2019/03/019}{\emph{JCAP} {\bfseries
  03} (2019) 019} [\href{https://arxiv.org/abs/1811.05988}{{\ttfamily
  1811.05988}}].

\bibitem{Ishiwata:2019aet}
K.~Ishiwata, O.~Macias, S.~Ando and M.~Arimoto, \emph{{Probing heavy dark
  matter decays with multi-messenger astrophysical data}},
  \href{https://doi.org/10.1088/1475-7516/2020/01/003}{\emph{JCAP} {\bfseries
  01} (2020) 003} [\href{https://arxiv.org/abs/1907.11671}{{\ttfamily
  1907.11671}}].

\bibitem{Esmaili:2021yaw}
A.~Esmaili and P.D.~Serpico, \emph{{First implications of Tibet
  AS\ensuremath{\gamma} data for heavy dark matter}},
  \href{https://doi.org/10.1103/PhysRevD.104.L021301}{\emph{Phys. Rev. D}
  {\bfseries 104} (2021) L021301}
  [\href{https://arxiv.org/abs/2105.01826}{{\ttfamily 2105.01826}}].

\bibitem{Maity:2021umk}
T.N.~Maity, A.K.~Saha, A.~Dubey and R.~Laha, \emph{{Search for dark matter
  using sub-PeV \ensuremath{\gamma}-rays observed by Tibet
  AS\ensuremath{\gamma}}},
  \href{https://doi.org/10.1103/PhysRevD.105.L041301}{\emph{Phys. Rev. D}
  {\bfseries 105} (2022) L041301}
  [\href{https://arxiv.org/abs/2105.05680}{{\ttfamily 2105.05680}}].

\bibitem{Chianese:2021jke}
M.~Chianese, D.F.G.~Fiorillo, R.~Hajjar, G.~Miele and N.~Saviano,
  \emph{{Constraints on heavy decaying dark matter with current gamma-ray
  measurements}},
  \href{https://doi.org/10.1088/1475-7516/2021/11/035}{\emph{JCAP} {\bfseries
  11} (2021) 035} [\href{https://arxiv.org/abs/2108.01678}{{\ttfamily
  2108.01678}}].

\bibitem{Dominguez:2010bv}
A.~Dominguez et~al., \emph{{Extragalactic Background Light Inferred from AEGIS
  Galaxy SED-type Fractions}},
  \href{https://doi.org/10.1111/j.1365-2966.2010.17631.x}{\emph{Mon. Not. Roy.
  Astron. Soc.} {\bfseries 410} (2011) 2556}
  [\href{https://arxiv.org/abs/1007.1459}{{\ttfamily 1007.1459}}].

\bibitem{Stecker:2016fsg}
F.W.~Stecker, S.T.~Scully and M.A.~Malkan, \emph{{An Empirical Determination of
  the Intergalactic Background Light from UV to FIR Wavelengths Using FIR Deep
  Galaxy Surveys and the Gamma-ray Opacity of the Universe}},
  \href{https://doi.org/10.3847/0004-637X/827/1/6}{\emph{Astrophys. J.}
  {\bfseries 827} (2016) 6} [\href{https://arxiv.org/abs/1605.01382}{{\ttfamily
  1605.01382}}].

\bibitem{Finke:2009xi}
J.D.~Finke, S.~Razzaque and C.D.~Dermer, \emph{{Modeling the Extragalactic
  Background Light from Stars and Dust}},
  \href{https://doi.org/10.1088/0004-637X/712/1/238}{\emph{Astrophys. J.}
  {\bfseries 712} (2010) 238}
  [\href{https://arxiv.org/abs/0905.1115}{{\ttfamily 0905.1115}}].

\bibitem{Kneiske:2010pt}
T.M.~Kneiske and H.~Dole, \emph{{A Lower-Limit Flux for the Extragalactic
  Background Light}},
  \href{https://doi.org/10.1051/0004-6361/200912000}{\emph{Astron. Astrophys.}
  {\bfseries 515} (2010) A19}
  [\href{https://arxiv.org/abs/1001.2132}{{\ttfamily 1001.2132}}].

\bibitem{Gilmore2012}
R.C.~{Gilmore}, R.S.~{Somerville}, J.R.~{Primack} and A.~{Dom{\'\i}nguez},
  \emph{{Semi-analytic modelling of the extragalactic background light and
  consequences for extragalactic gamma-ray spectra}},
  \href{https://doi.org/10.1111/j.1365-2966.2012.20841.x}{\emph{Monthly Notices
  of the RAS} {\bfseries 422} (2012) 3189}
  [\href{https://arxiv.org/abs/1104.0671}{{\ttfamily 1104.0671}}].

\bibitem{Helgason:2012xj}
K.~Helgason and A.~Kashlinsky, \emph{{Reconstructing the
  \textbackslash{}gamma-ray Photon Optical Depth of the Universe to
  z\textasciitilde{}4 from Multiwavelength Galaxy Survey Data}},
  \href{https://doi.org/10.1088/2041-8205/758/1/L13}{\emph{Astrophys. J. Lett.}
  {\bfseries 758} (2012) L13}
  [\href{https://arxiv.org/abs/1208.4364}{{\ttfamily 1208.4364}}].

\bibitem{Inoue:2012bk}
Y.~Inoue, S.~Inoue, M.A.R.~Kobayashi, R.~Makiya, Y.~Niino and T.~Totani,
  \emph{{Extragalactic Background Light from Hierarchical Galaxy Formation:
  Gamma-ray Attenuation up to the Epoch of Cosmic Reionization and the First
  Stars}}, \href{https://doi.org/10.1088/0004-637X/768/2/197}{\emph{Astrophys.
  J.} {\bfseries 768} (2013) 197}
  [\href{https://arxiv.org/abs/1212.1683}{{\ttfamily 1212.1683}}].

\bibitem{Franceschini2017}
A.~{Franceschini} and G.~{Rodighiero}, \emph{{The extragalactic background
  light revisited and the cosmic photon-photon opacity}},
  \href{https://doi.org/10.1051/0004-6361/201629684}{\emph{Astronomy and
  Astrophysics} {\bfseries 603} (2017) A34}
  [\href{https://arxiv.org/abs/1705.10256}{{\ttfamily 1705.10256}}].

\bibitem{Andrews:2017ima}
S.K.~Andrews, S.P.~Driver, L.J.~Davies, C.d.P.~Lagos and A.S.G.~Robotham,
  \emph{{Modelling the cosmic spectral energy distribution and extragalactic
  background light over all time}},
  \href{https://doi.org/10.1093/mnras/stx2843}{\emph{Mon. Not. Roy. Astron.
  Soc.} {\bfseries 474} (2018) 898}
  [\href{https://arxiv.org/abs/1710.11329}{{\ttfamily 1710.11329}}].

\bibitem{TibetASgamma:2021tpz}
{\scshape Tibet ASgamma} collaboration, \emph{{First Detection of sub-PeV
  Diffuse Gamma Rays from the Galactic Disk: Evidence for Ubiquitous Galactic
  Cosmic Rays beyond PeV Energies}},
  \href{https://doi.org/10.1103/PhysRevLett.126.141101}{\emph{Phys. Rev. Lett.}
  {\bfseries 126} (2021) 141101}
  [\href{https://arxiv.org/abs/2104.05181}{{\ttfamily 2104.05181}}].

\bibitem{Fermi-LAT:2014ryh}
{\scshape Fermi-LAT} collaboration, \emph{{The spectrum of isotropic diffuse
  gamma-ray emission between 100 MeV and 820 GeV}}, {\emph{Astrophys. J.}
  {\bfseries 799} (2015) 86} [\href{https://arxiv.org/abs/1410.3696}{{\ttfamily
  1410.3696}}].

\bibitem{Benito_2021}
M.~Benito, F.~Iocco and A.~Cuoco, \emph{Uncertainties in the galactic dark
  matter distribution: An update},
  \href{https://doi.org/10.1016/j.dark.2021.100826}{\emph{Physics of the Dark
  Universe} {\bfseries 32} (2021) 100826}.

\bibitem{Planck:2018vyg}
{\scshape Planck} collaboration, \emph{{Planck 2018 results. VI. Cosmological
  parameters}},
  \href{https://doi.org/10.1051/0004-6361/201833910}{\emph{Astron. Astrophys.}
  {\bfseries 641} (2020) A6}
  [\href{https://arxiv.org/abs/1807.06209}{{\ttfamily 1807.06209}}].

\bibitem{Berezinsky:1975zz}
V.S.~Berezinsky and A.Y.~Smirnov, \emph{{Cosmic neutrinos of ultra-high
  energies and detection possibility}},
  \href{https://doi.org/10.1007/BF00643157}{\emph{Astrophys. Space Sci.}
  {\bfseries 32} (1975) 461}.

\bibitem{Coppi:1996ze}
P.S.~Coppi and F.A.~Aharonian, \emph{{Constraints on the VHE emissivity of the
  universe from the diffuse GeV gamma-ray background}},
  \href{https://doi.org/10.1086/310883}{\emph{Astrophys. J. Lett.} {\bfseries
  487} (1997) L9} [\href{https://arxiv.org/abs/astro-ph/9610176}{{\ttfamily
  astro-ph/9610176}}].

\bibitem{Venters:2010bq}
T.M.~Venters, \emph{{Contribution to the Extragalactic Gamma-ray Background
  from the Cascades of Very-high Energy Gamma Rays from Blazars}},
  \href{https://doi.org/10.1088/0004-637X/710/2/1530}{\emph{Astrophys. J.}
  {\bfseries 710} (2010) 1530}
  [\href{https://arxiv.org/abs/1001.1363}{{\ttfamily 1001.1363}}].

\bibitem{Murase:2012df}
K.~Murase, J.F.~Beacom and H.~Takami, \emph{{Gamma-Ray and Neutrino Backgrounds
  as Probes of the High-Energy Universe: Hints of Cascades, General
  Constraints, and Implications for TeV Searches}},
  \href{https://doi.org/10.1088/1475-7516/2012/08/030}{\emph{JCAP} {\bfseries
  08} (2012) 030} [\href{https://arxiv.org/abs/1205.5755}{{\ttfamily
  1205.5755}}].

\bibitem{Berezinsky:2016feh}
V.~Berezinsky and O.~Kalashev, \emph{{High energy electromagnetic cascades in
  extragalactic space: physics and features}},
  \href{https://doi.org/10.1103/PhysRevD.94.023007}{\emph{Phys. Rev. D}
  {\bfseries 94} (2016) 023007}
  [\href{https://arxiv.org/abs/1603.03989}{{\ttfamily 1603.03989}}].

\bibitem{PhysRevD.58.043004}
S.~Lee, \emph{Propagation of extragalactic high energy cosmic and
  \ensuremath{\gamma} rays},
  \href{https://doi.org/10.1103/PhysRevD.58.043004}{\emph{Phys. Rev. D}
  {\bfseries 58} (1998) 043004}.

\bibitem{Vladimirov:2010aq}
A.E.~Vladimirov, S.W.~Digel, G.~Johannesson, P.F.~Michelson, I.V.~Moskalenko,
  P.L.~Nolan et~al., \emph{{GALPROP WebRun: an internet-based service for
  calculating galactic cosmic ray propagation and associated photon
  emissions}}, \href{https://doi.org/10.1016/j.cpc.2011.01.017}{\emph{Comput.
  Phys. Commun.} {\bfseries 182} (2011) 1156}
  [\href{https://arxiv.org/abs/1008.3642}{{\ttfamily 1008.3642}}].

\bibitem{Kachelriess:2011bi}
M.~Kachelriess, S.~Ostapchenko and R.~Tomas, \emph{{ELMAG: A Monte Carlo
  simulation of electromagnetic cascades on the extragalactic background light
  and in magnetic fields}},
  \href{https://doi.org/10.1016/j.cpc.2011.12.025}{\emph{Comput. Phys. Commun.}
  {\bfseries 183} (2012) 1036}
  [\href{https://arxiv.org/abs/1106.5508}{{\ttfamily 1106.5508}}].

\bibitem{Kalashev:2014xna}
O.E.~Kalashev and E.~Kido, \emph{{Simulations of Ultra High Energy Cosmic Rays
  propagation}}, \href{https://doi.org/10.1134/S1063776115040056}{\emph{J. Exp.
  Theor. Phys.} {\bfseries 120} (2015) 790}
  [\href{https://arxiv.org/abs/1406.0735}{{\ttfamily 1406.0735}}].

\bibitem{Blanco:2018bbf}
C.~Blanco, \emph{{$\gamma$-cascade: a simple program to compute cosmological
  gamma-ray propagation}},
  \href{https://doi.org/10.1088/1475-7516/2019/01/013}{\emph{JCAP} {\bfseries
  01} (2019) 013} [\href{https://arxiv.org/abs/1804.00005}{{\ttfamily
  1804.00005}}].

\bibitem{AlvesBatista:2016vpy}
R.~Alves~Batista, A.~Dundovic, M.~Erdmann, K.-H.~Kampert, D.~Kuempel,
  G.~M\"uller et~al., \emph{{CRPropa 3 - a Public Astrophysical Simulation
  Framework for Propagating Extraterrestrial Ultra-High Energy Particles}},
  \href{https://doi.org/10.1088/1475-7516/2016/05/038}{\emph{JCAP} {\bfseries
  05} (2016) 038} [\href{https://arxiv.org/abs/1603.07142}{{\ttfamily
  1603.07142}}].

\bibitem{Heiter:2017cev}
C.~Heiter, D.~Kuempel, D.~Walz and M.~Erdmann, \emph{{Production and
  propagation of ultra-high energy photons using CRPropa 3}},
  \href{https://doi.org/10.1016/j.astropartphys.2018.05.003}{\emph{Astropart.
  Phys.} {\bfseries 102} (2018) 39}
  [\href{https://arxiv.org/abs/1710.11406}{{\ttfamily 1710.11406}}].

\bibitem{Driver:2016krv}
S.P.~Driver, S.K.~Andrews, L.J.~Davies, A.S.G.~Robotham, A.H.~Wright,
  R.A.~Windhorst et~al., \emph{{Measurements of Extragalactic Background Light
  From the far UV to the far IR From Deep Ground- and Space-based Galaxy
  Counts}}, \href{https://doi.org/10.3847/0004-637X/827/2/108}{\emph{Astrophys.
  J.} {\bfseries 827} (2016) 108}
  [\href{https://arxiv.org/abs/1605.01523}{{\ttfamily 1605.01523}}].

\bibitem{Fermi-LAT:2018lqt}
{\scshape Fermi-LAT} collaboration, \emph{{A gamma-ray determination of the
  Universe\textquoteright{}s star formation history}},
  \href{https://doi.org/10.1126/science.aat8123}{\emph{Science} {\bfseries 362}
  (2018) 1031} [\href{https://arxiv.org/abs/1812.01031}{{\ttfamily
  1812.01031}}].

\bibitem{Singh:2020kmg}
K.K.~Singh and P.J.~Meintjes, \emph{{Extragalactic background light models and
  GeV-TeV observation of blazars}},
  \href{https://arxiv.org/abs/2004.01933}{{\ttfamily 2004.01933}}.

\bibitem{Saveliev:2020ynu}
A.~Saveliev and R.~Alves~Batista, \emph{{The intrinsic gamma-ray spectrum of
  TXS 0506+056: intergalactic propagation effects}},
  \href{https://doi.org/10.1093/mnras/staa3403}{\emph{Mon. Not. Roy. Astron.
  Soc.} {\bfseries 500} (2020) 2188}
  [\href{https://arxiv.org/abs/2009.09772}{{\ttfamily 2009.09772}}].

\bibitem{Dwek_2013}
E.~Dwek and F.~Krennrich, \emph{The extragalactic background light and the
  gamma-ray opacity of the universe},
  \href{https://doi.org/10.1016/j.astropartphys.2012.09.003}{\emph{Astroparticle
  Physics} {\bfseries 43} (2013) 112}.

\bibitem{Biteau_2015}
J.~Biteau and D.A.~Williams, \emph{{THE} {EXTRAGALACTIC} {BACKGROUND} {LIGHT},
  {THE} {HUBBLE} {CONSTANT}, {AND} {ANOMALIES}: {CONCLUSIONS} {FROM} 20 {YEARS}
  {OF} {TeV} {GAMMA}-{RAY} {OBSERVATIONS}},
  \href{https://doi.org/10.1088/0004-637x/812/1/60}{\emph{The Astrophysical
  Journal} {\bfseries 812} (2015) 60}.

\bibitem{Abeysekara_2019}
A.U.~Abeysekara, A.~Archer, W.~Benbow, R.~Bird, A.~Brill, R.~Brose et~al.,
  \emph{Measurement of the extragalactic background light spectral energy
  distribution with {VERITAS}},
  \href{https://doi.org/10.3847/1538-4357/ab4817}{\emph{The Astrophysical
  Journal} {\bfseries 885} (2019) 150}.

\bibitem{Hess2017}
and H.~Abdalla, A.~Abramowski, F.~Aharonian, F.A.~Benkhali, A.G.~Akhperjanian,
  T.~Andersson et~al., \emph{Measurement of the {EBL} spectral energy
  distribution using the {VHE} $\gamma$-ray spectra of h.e.s.s. blazars},
  \href{https://doi.org/10.1051/0004-6361/201731200}{\emph{Astronomy
  Astrophysics} {\bfseries 606} (2017) A59}.

\bibitem{Arguelles:2019ouk}
C.A.~Arg\"uelles, A.~Diaz, A.~Kheirandish, A.~Olivares-Del-Campo, I.~Safa and
  A.C.~Vincent, \emph{{Dark matter annihilation to neutrinos}},
  \href{https://doi.org/10.1103/RevModPhys.93.035007}{\emph{Rev. Mod. Phys.}
  {\bfseries 93} (2021) 035007}
  [\href{https://arxiv.org/abs/1912.09486}{{\ttfamily 1912.09486}}].

\bibitem{Sjostrand:2014zea}
T.~Sj\"ostrand, S.~Ask, J.R.~Christiansen, R.~Corke, N.~Desai, P.~Ilten et~al.,
  \emph{{An introduction to PYTHIA 8.2}},
  \href{https://doi.org/10.1016/j.cpc.2015.01.024}{\emph{Comput. Phys. Commun.}
  {\bfseries 191} (2015) 159}
  [\href{https://arxiv.org/abs/1410.3012}{{\ttfamily 1410.3012}}].

\bibitem{Cirelli:2010xx}
M.~Cirelli, G.~Corcella, A.~Hektor, G.~Hutsi, M.~Kadastik, P.~Panci et~al.,
  \emph{{PPPC 4 DM ID: A Poor Particle Physicist Cookbook for Dark Matter
  Indirect Detection}},
  \href{https://doi.org/10.1088/1475-7516/2012/10/E01}{\emph{JCAP} {\bfseries
  03} (2011) 051} [\href{https://arxiv.org/abs/1012.4515}{{\ttfamily
  1012.4515}}].

\bibitem{Jansson_2012}
R.~Jansson and G.R.~Farrar, \emph{A {NEW} {MODEL} {OF} {THE} {GALACTIC}
  {MAGNETIC} {FIELD}},
  \href{https://doi.org/10.1088/0004-637x/757/1/14}{\emph{The Astrophysical
  Journal} {\bfseries 757} (2012) 14}.

\end{thebibliography}\endgroup


\appendix

\section{Simulation Details\label{simulation}}

To propagate photons, electrons, and positrons over galactic and extragalactic scales, we use \texttt{CRPropa}~\cite{AlvesBatista:2016vpy}, a tool developed to simulate ultra-high-energy nuclei within galactic and extragalactic environments. For the propagations that we simulate in \texttt{CRPropa}, we enable the following processes: pair production, double pair production, triplet pair production, Inverse-Compton scattering, and synchrotron radiation. For galactic propagation, we set the magnetic field to be the JF12 field model~\cite{Jansson_2012}, while for extragalactic propagation we assume no additional magnetic field contribution. In both cases, we inject an initial power-law spectrum that follows $\phi \sim (E/E_c)^{-1}$ and which we define uniformly over a solid sphere. For the galactic case, propagation takes places over distances of  0 to 100 kpc, while for the extragalactic case, we propagation between 100 kpc and 10 Gpc. \texttt{CRPropa} creates this initial spectrum via Monte Carlo generation, where we set the number of Monte Carlo simulations to be roughly $1.5\times 10^6$ per injected dark matter mass. Regarding the injected spectrum, the minimum and maximum $(E_c)$ energies vary with the dark matter mass that we consider such that $E_c = m_{\rm DM}/2$ and $E_{\rm min} = (m_{\rm DM}/2)\times 10^{-6}$. 

To then obtain the expected dark-matter decay spectrum from the \texttt{CRPropa} output following propagation, we apply a reweighting procedure where we group the photons that arrive at Earth according to their corresponding, unique Monte Carlo generated events. The reweighting scheme then assigns a new weight to each of these events that depends on that event's initial energy and position, such that the new weights correspond to a dark matter distribution. Schematically, the end result of this is that the our reweighting scheme takes the initial power-law distribution that is fed into \texttt{CRPropa} and returns the expected final, propagated spectrum that results from dark-matter decay for a given dark matter mass, lifetime, and decay channel. Although we apply this to the case of dark matter decay, this procedure could be generalized for any arbitrary, well-behaved distribution that needs to be propagated over some distance.

The weight that we assign to each event can be decomposed into a position weight, and energy weight, and a multiplicative prefactor that depends on dark matter parameters and that converts the result into units of flux. 

\subsection{Position Reweighting}

The position reweighting that we apply to each Monte Carlo event assigns a multiplicative factor replaces the Monte Carlos generation weights for a uniformly distributed source with those corresponding to the desired dark matter profile. Schematically, these position weights can be expressed as 
\begin{equation}
    w_r = \frac{w_{\rm physical}}{w_{\rm generated}} = \Big(\rho_{\rm DM}(r)\Big)\Bigg(\frac{D(r)}{\int_{r_{\rm min}}^{r_{\rm max}}D(r)}\Bigg)^{-1} = \rho_{\rm DM}(r)(r_{\rm max}-r_{\rm min}) \,,
\end{equation}
where $D(r)$ is the initial distribution, $\rho_{\rm DM}(r)$ is the dark matter distribution, and the denominator of the quantity in parentheses corresponds to an overall normalization included in the generated weights, which is a uniform distribution for our simulations.
The physical position weights depend on the profile that we consider. For galactic reweighting, for example, we use the generalized NFW profile $\rho_{\rm DM}(r)$ defined in Eq.~\eqref{eq:profile}. For the extragalactic position reweighting, we use $\rho_{\rm cr}\Omega_{\rm DM}/H(z)$, where we set the parameters to the values mentioned in ~\cref{sec:dm}.

\subsection{Energy Reweighting}
In a similar way, the energy reweighting acts as a multiplicative factor that replaces the Monte Carlo simulation weights for the power-law spectrum $\phi(E) \sim E^{-1}$ generated by \texttt{CRPropa} and returns a physical weight for each event that corresponds to the dark matter decay spectrum $\phi_{\rm DM} \propto {\rm d}N/{\rm d}E$ computed by \texttt{HDMSpectra}. These energy weights can be expressed as 
\begin{equation}
    w_E = \frac{w_{\rm physical}}{w_{\rm generated}} = \Big(\phi_{\rm DM}(E)\Big)\Bigg(\frac{\phi(E)}{\int_{E_{\rm min}}^{E_{\rm max}}\phi(E)}\Bigg)^{-1} = \phi_{\rm DM}(E)(E\times\ln(E_{\rm max}/E_{\rm min})) \,.
\end{equation}
For the extragalactic energy reweighting, there is also a redshift in the energy of the initial spectrum which resulting in an additional factor of $(1+z)$.

\section{Spectra of Photons Arriving at Earth\label{app:gamma_detail}}

\begin{landscape}
\begin{figure}[t!]
\begin{center}
    \includegraphics[width = 0.24\columnwidth]{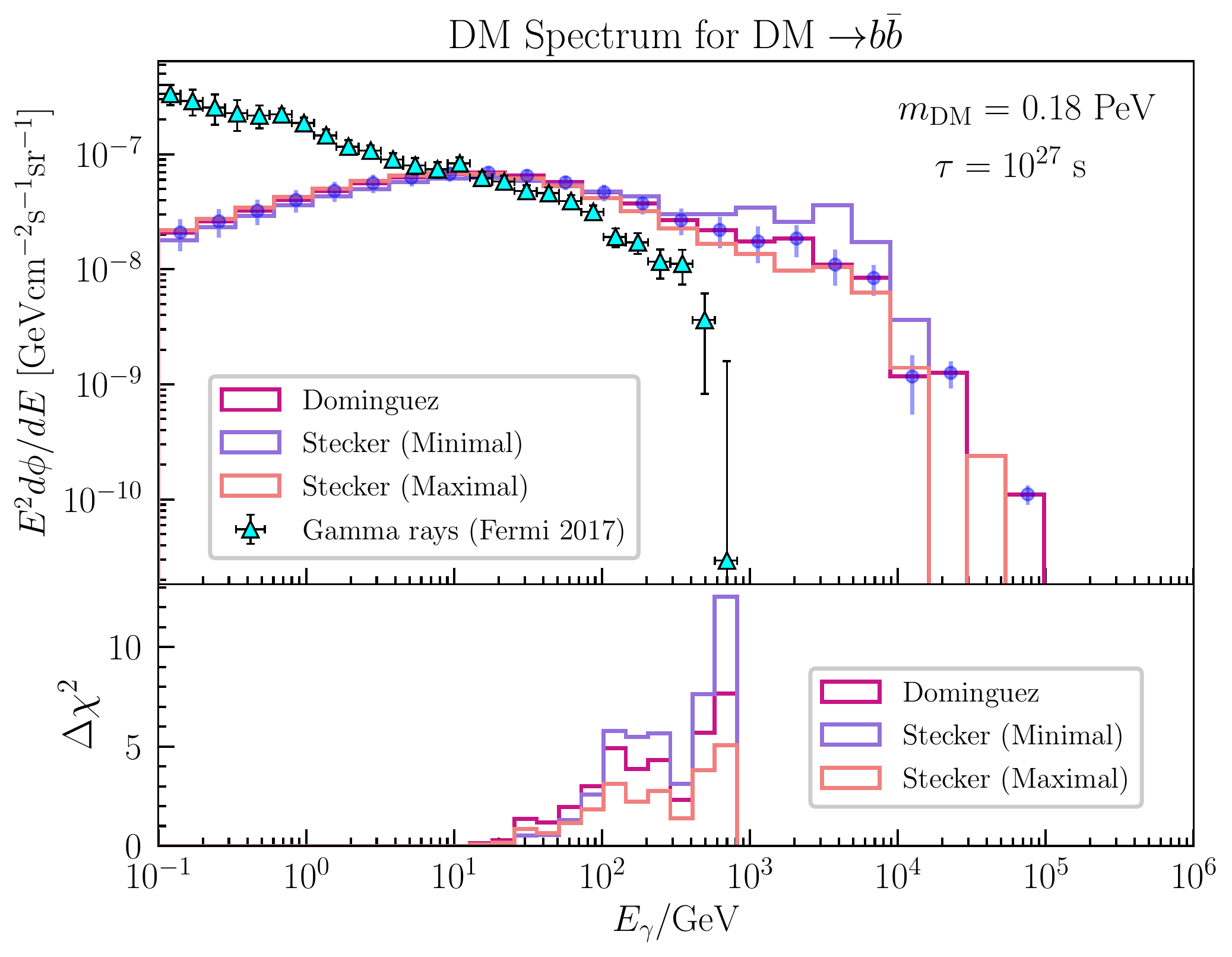}
    \includegraphics[width = 0.24\columnwidth]{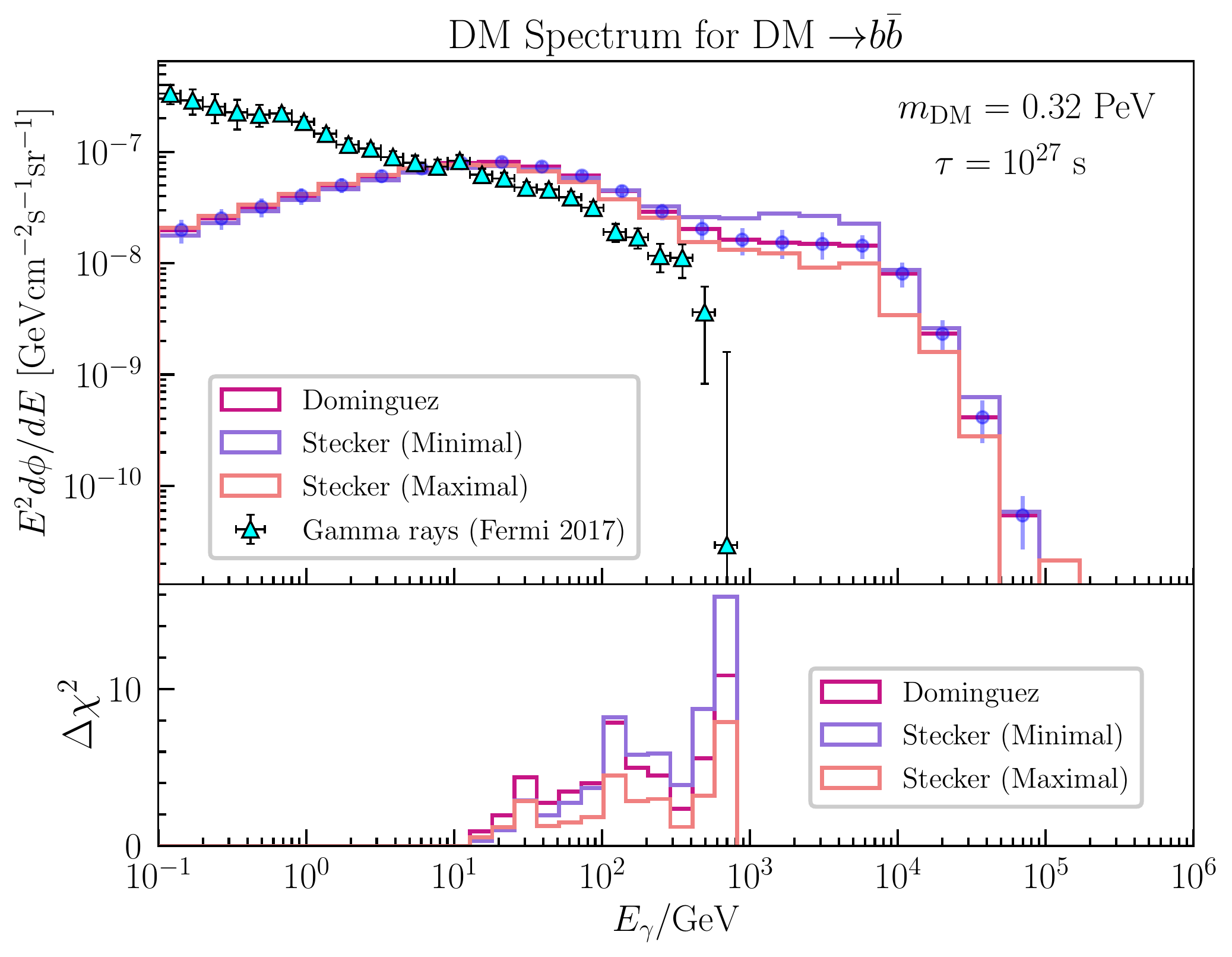}
    \includegraphics[width = 0.24\columnwidth]{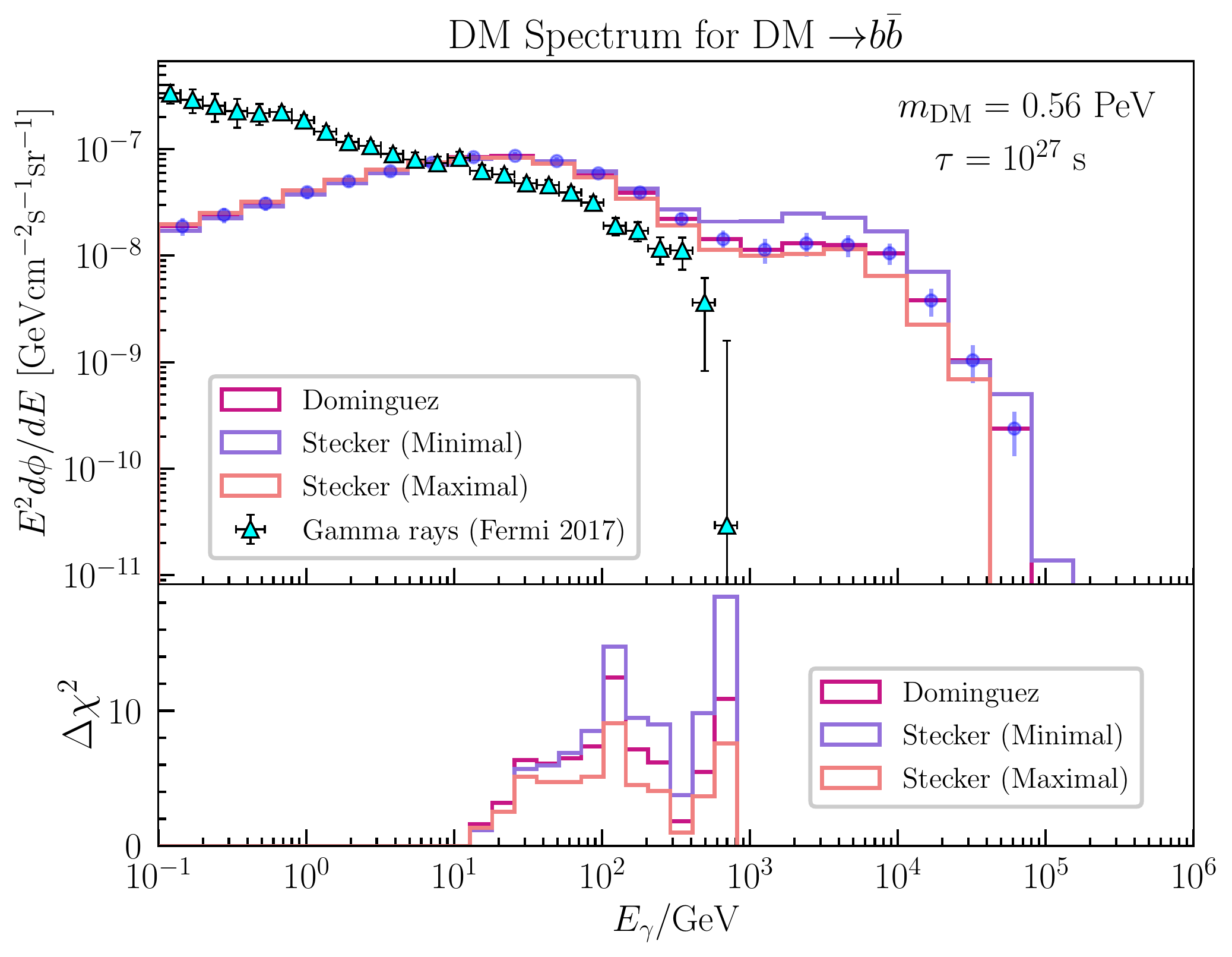}
    \includegraphics[width = 0.24\columnwidth]{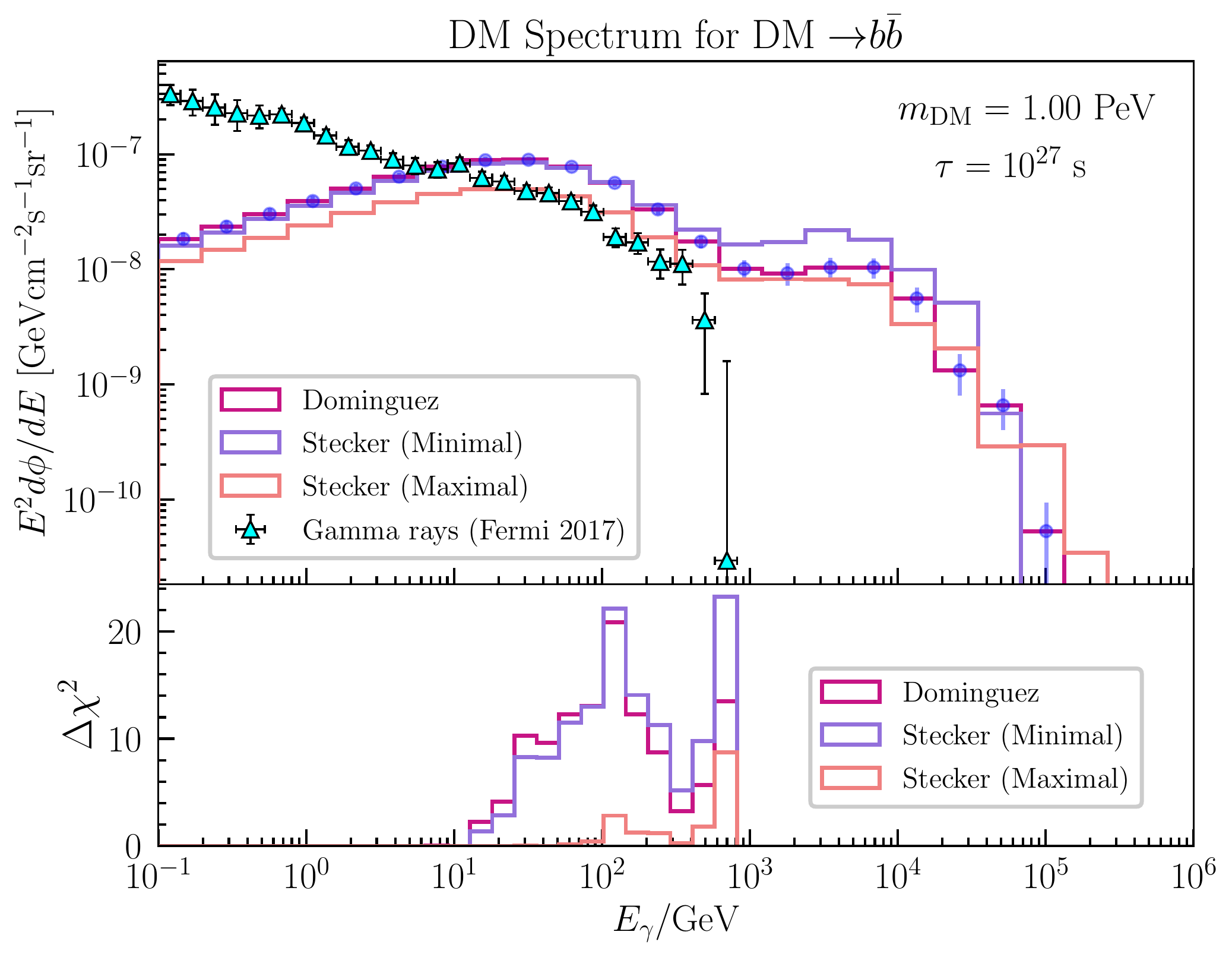}\\
    \includegraphics[width = 0.24\columnwidth]{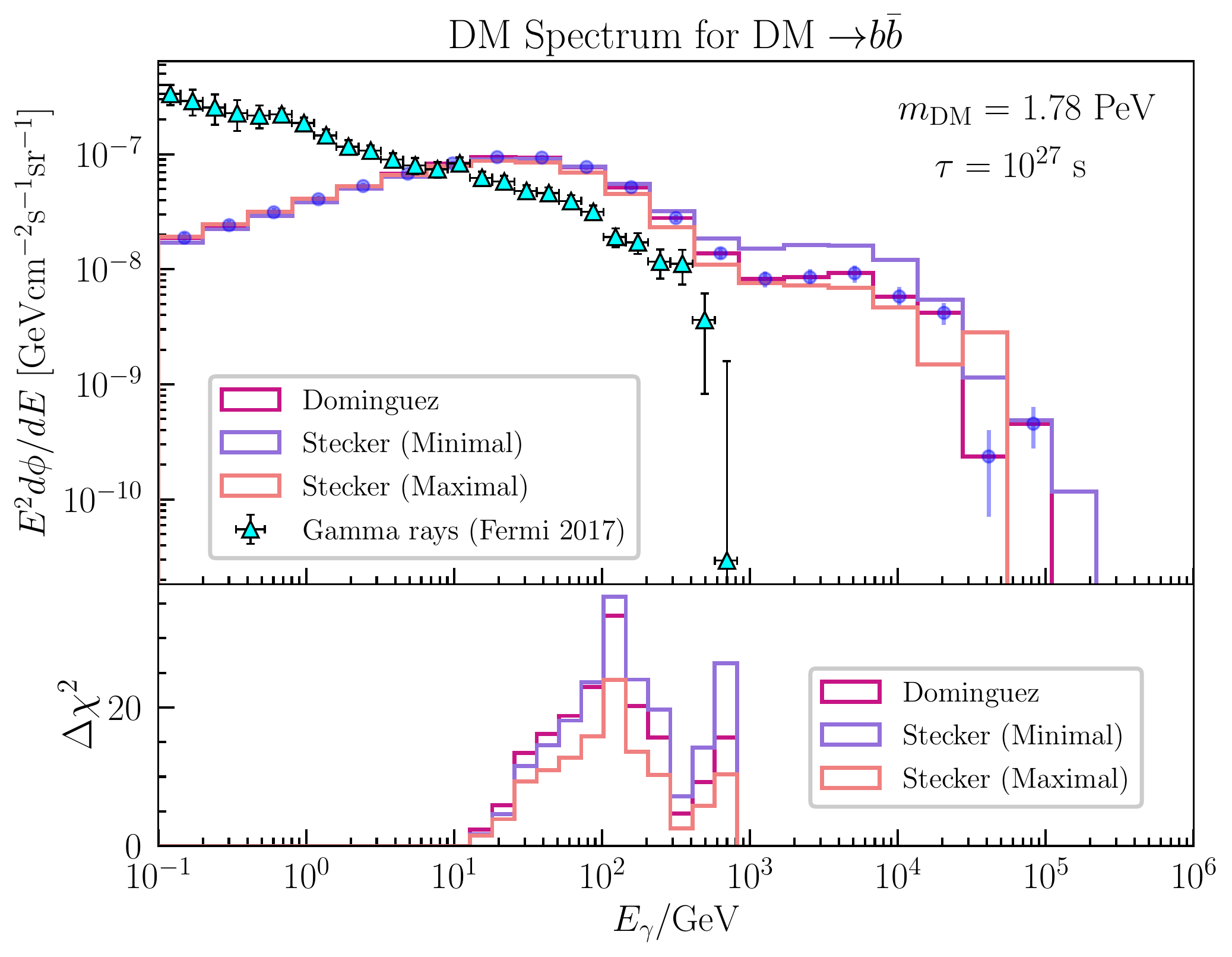}
    \includegraphics[width = 0.24\columnwidth]{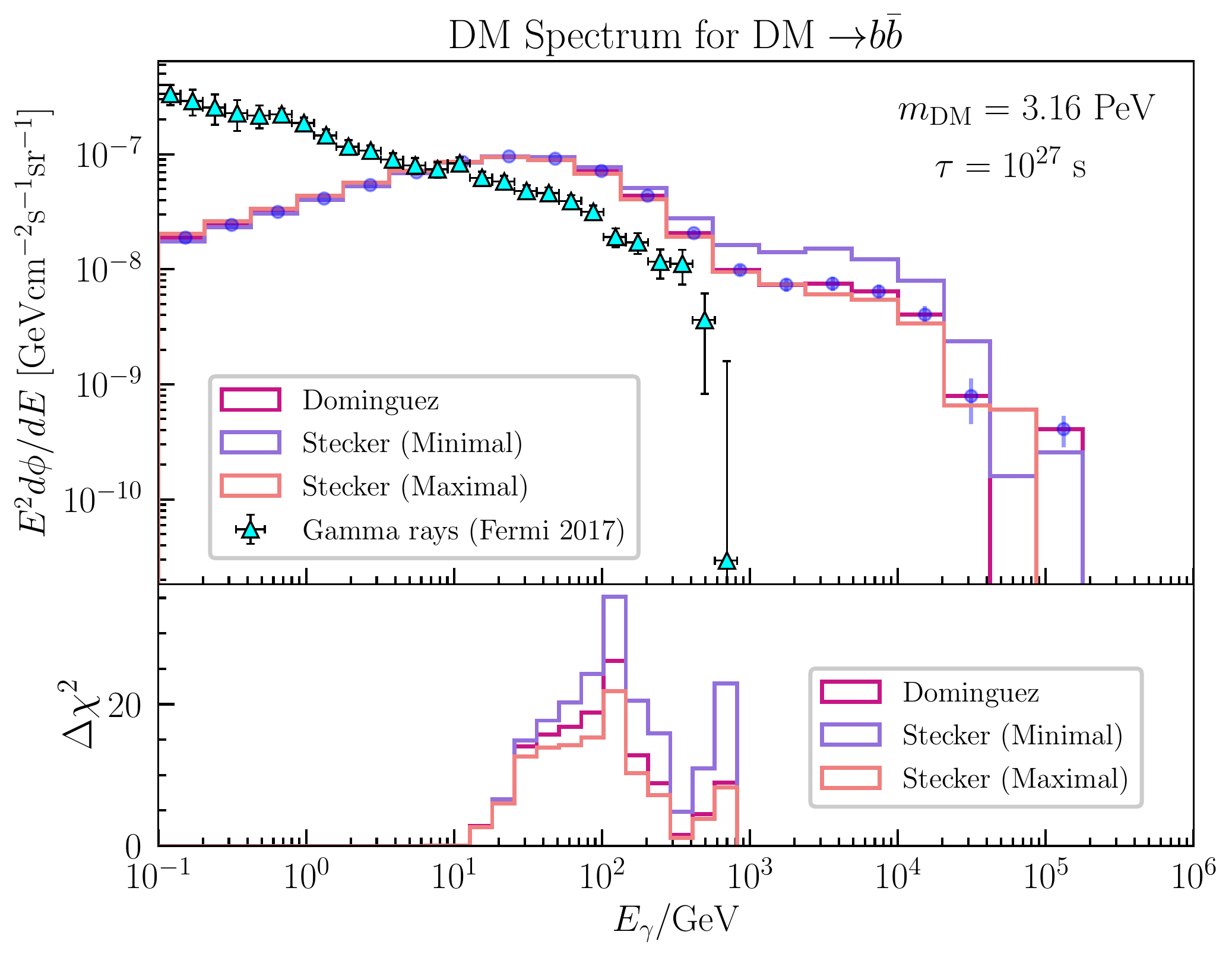}
    \includegraphics[width = 0.24\columnwidth]{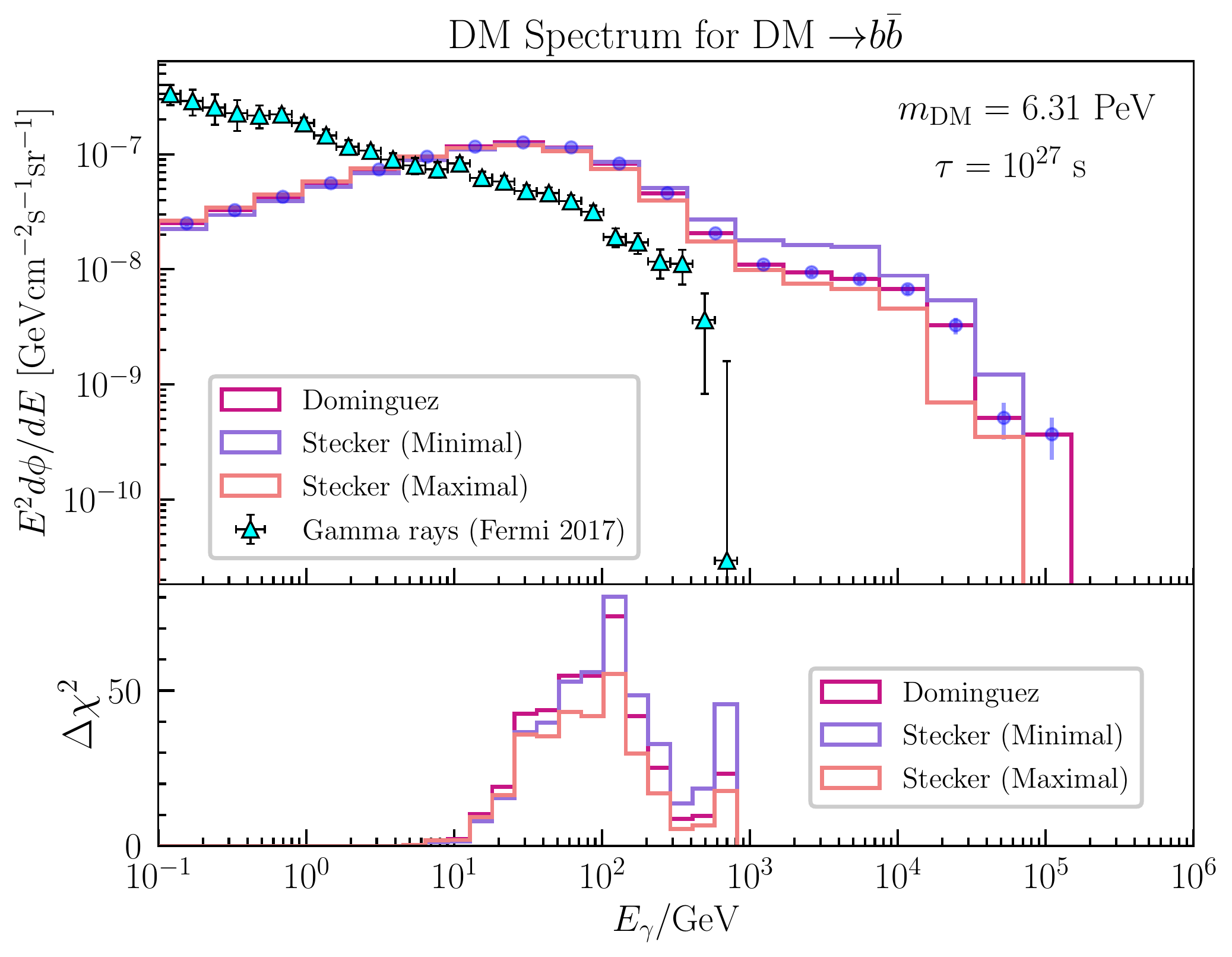}
    \includegraphics[width = 0.24\columnwidth]{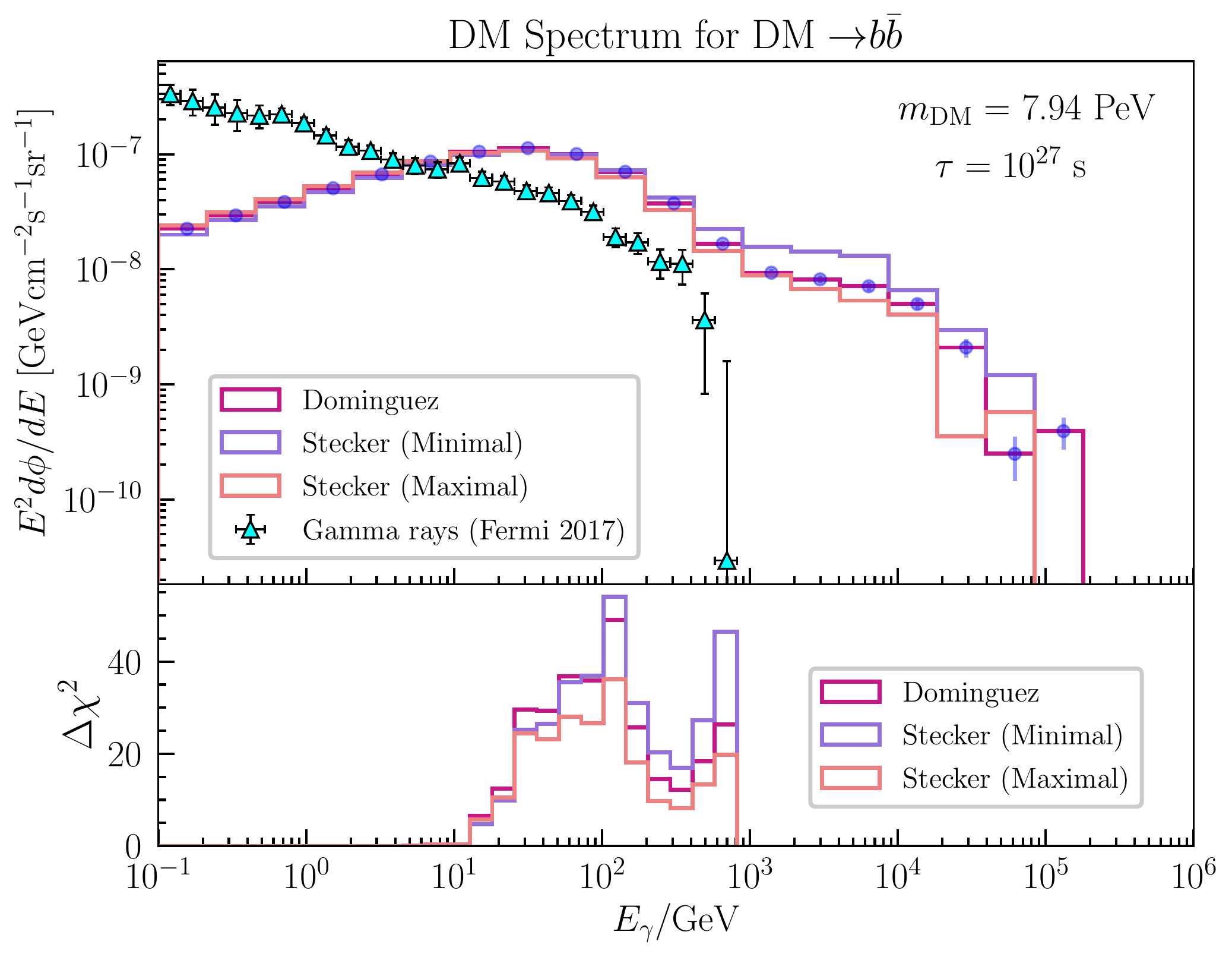}
\end{center}
    \caption{Final dark matter gamma-ray flux from the extragalactic secondary $\gamma \to \gamma$ component, for different dark matter masses at a lifetime of $10^{27}$ seconds. Different colors correspond to different EBL models considered for the propagation through the extragalactic space. The bottom panels show the $\Delta \chi^2$ calculated by comparing our Monte Carlo simulations to Fermi-LAT measurements of the IGRB spectrum, displayed by the triangular points. The circular data points show the mean prediction and the corresponding errors from the Monte Carlo simulations according to the Dominguez \textit{et al.} EBL model.}
    \label{fig:EG-Dark-Matter-Gamma}
\end{figure}
\end{landscape}

\begin{landscape}
\begin{figure}[t!]
\begin{center}
    \includegraphics[width = 0.24\columnwidth]{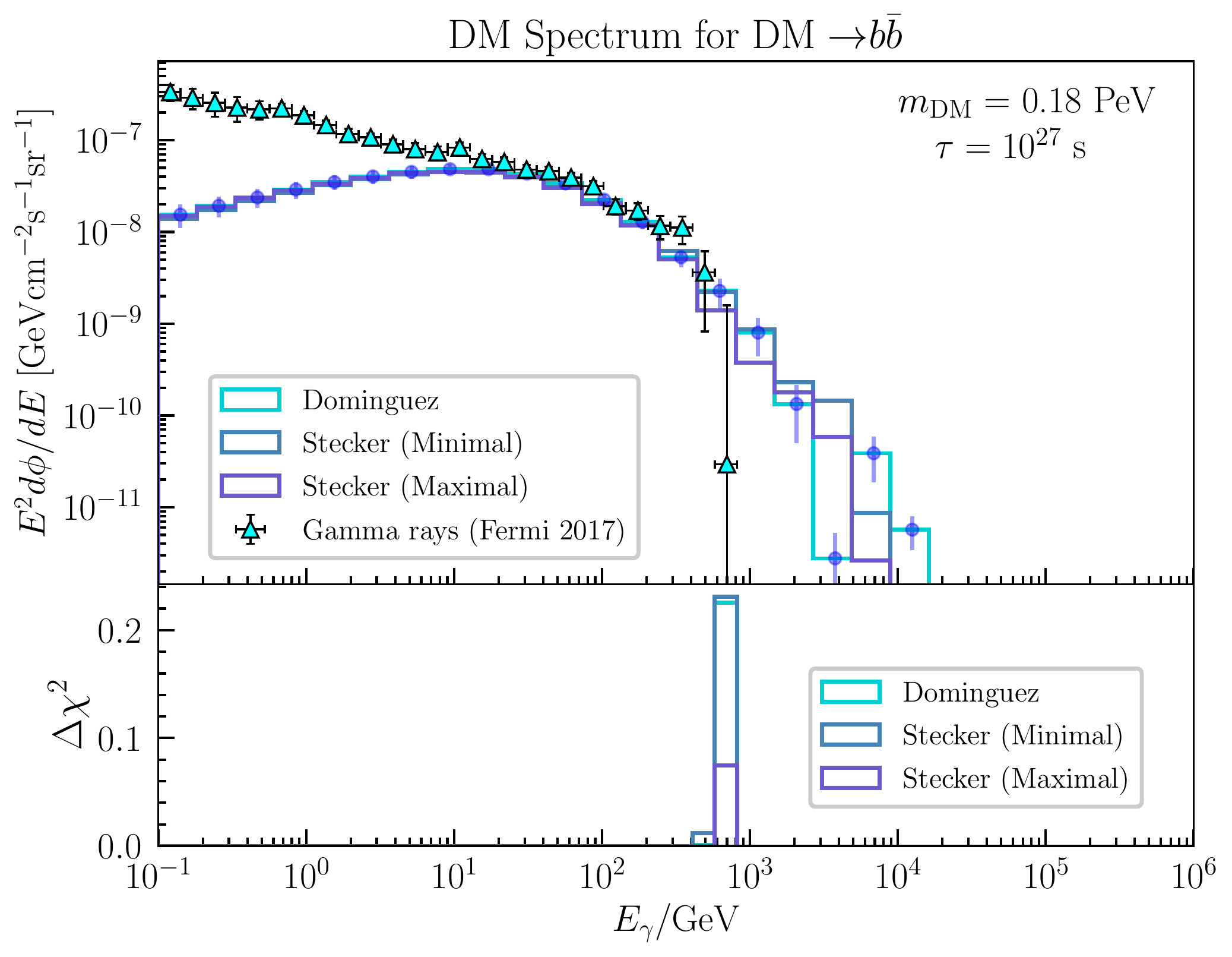}
    \includegraphics[width = 0.24\columnwidth]{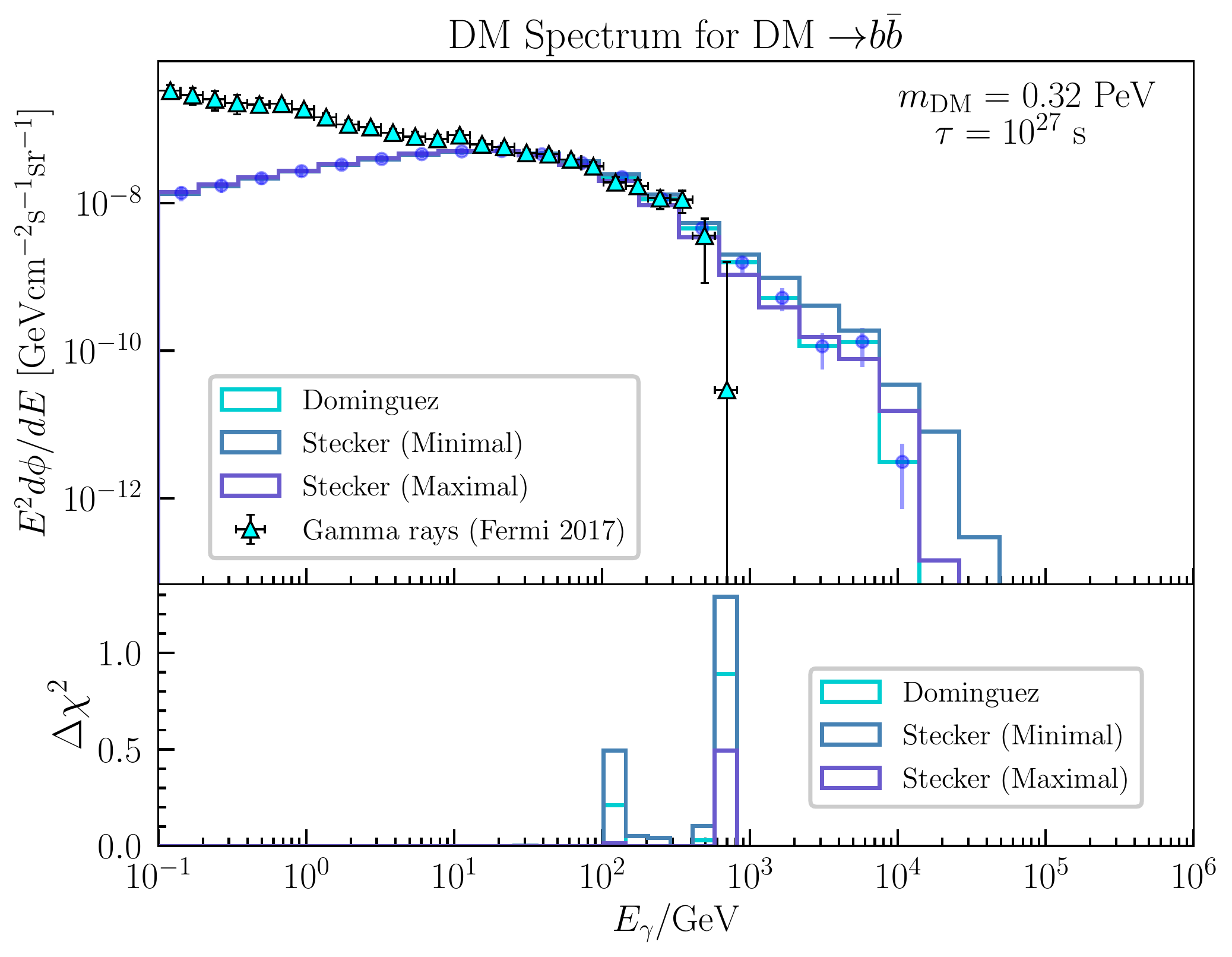}
    \includegraphics[width = 0.24\columnwidth]{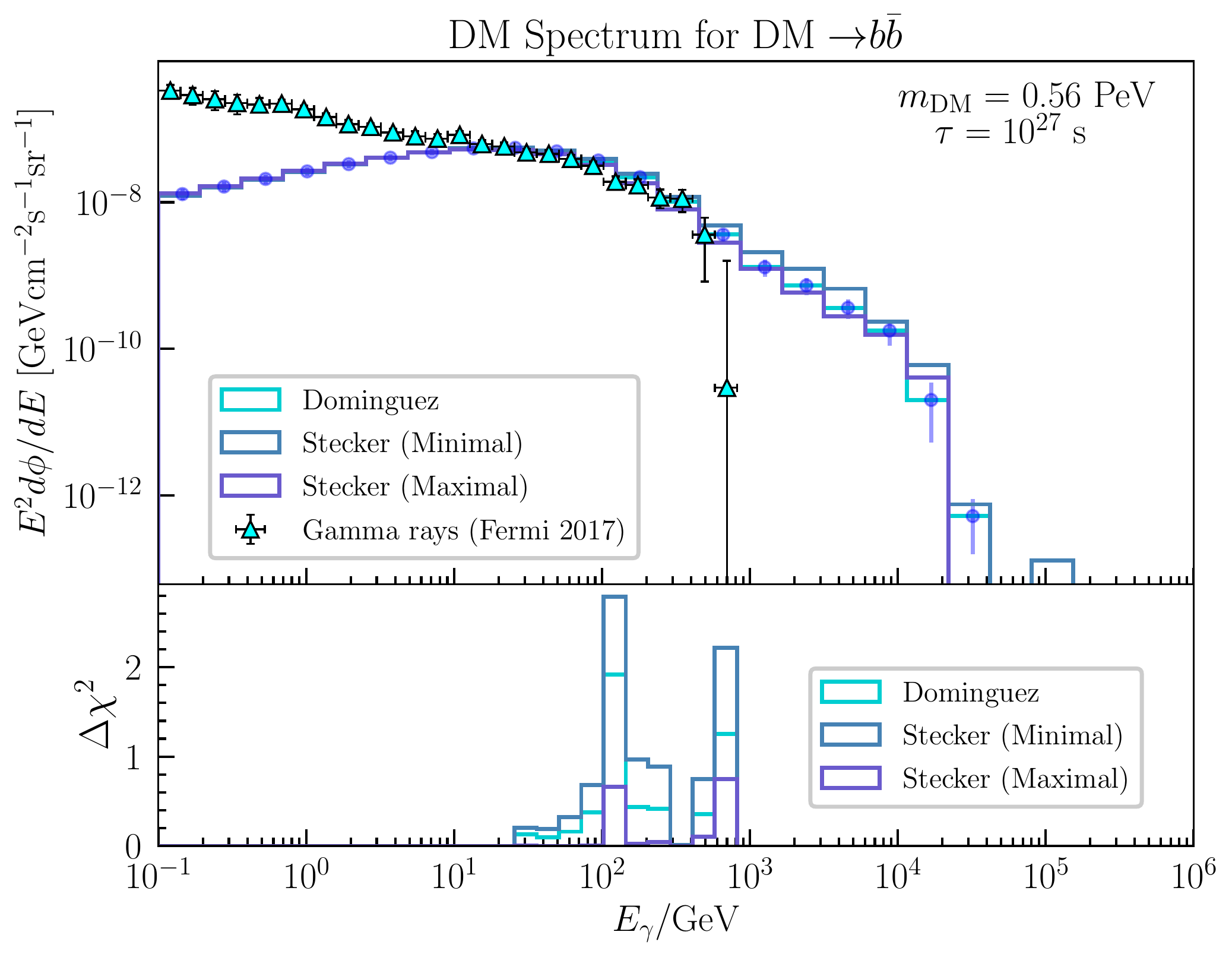}
    \includegraphics[width = 0.24\columnwidth]{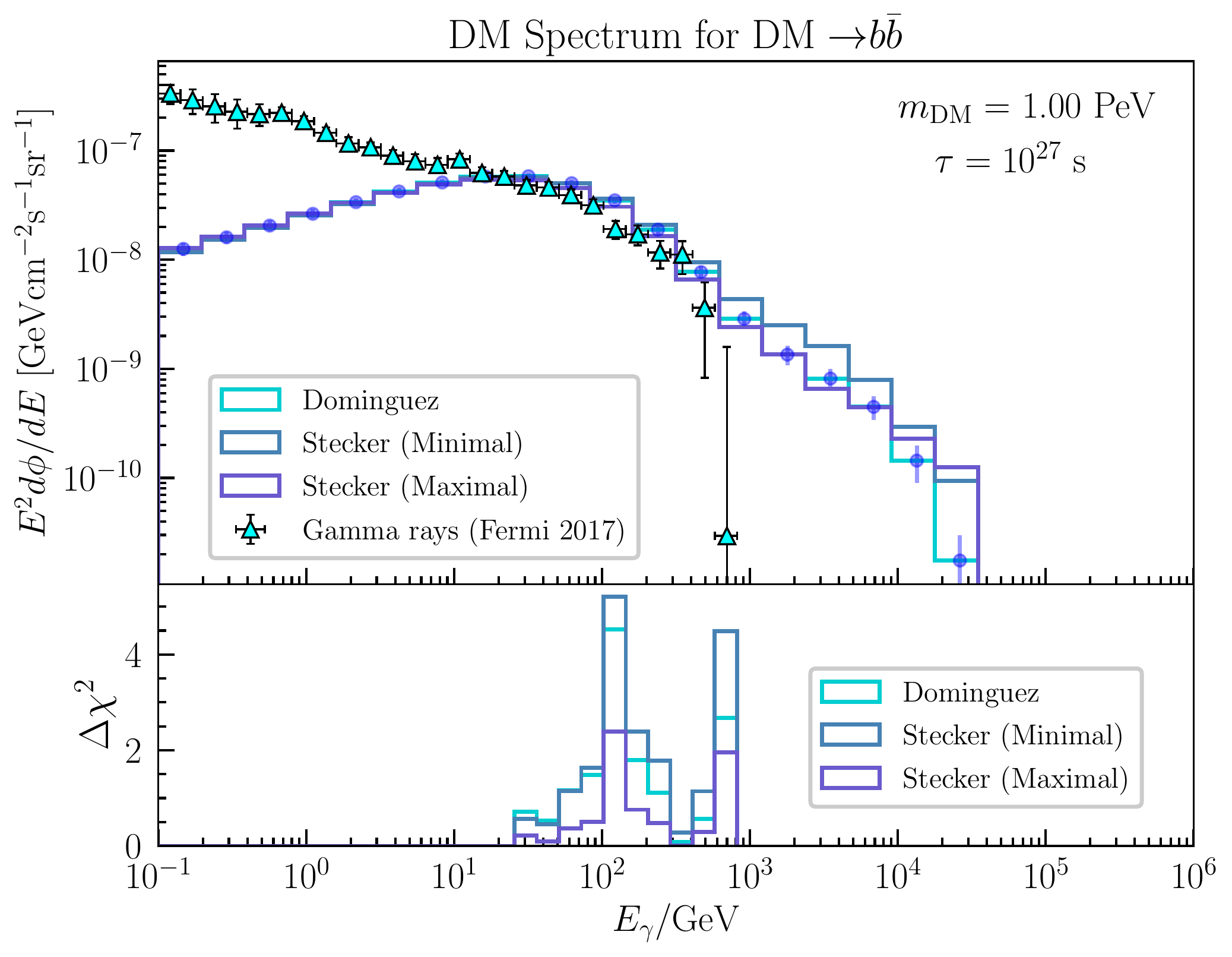}\\
    \includegraphics[width = 0.24\columnwidth]{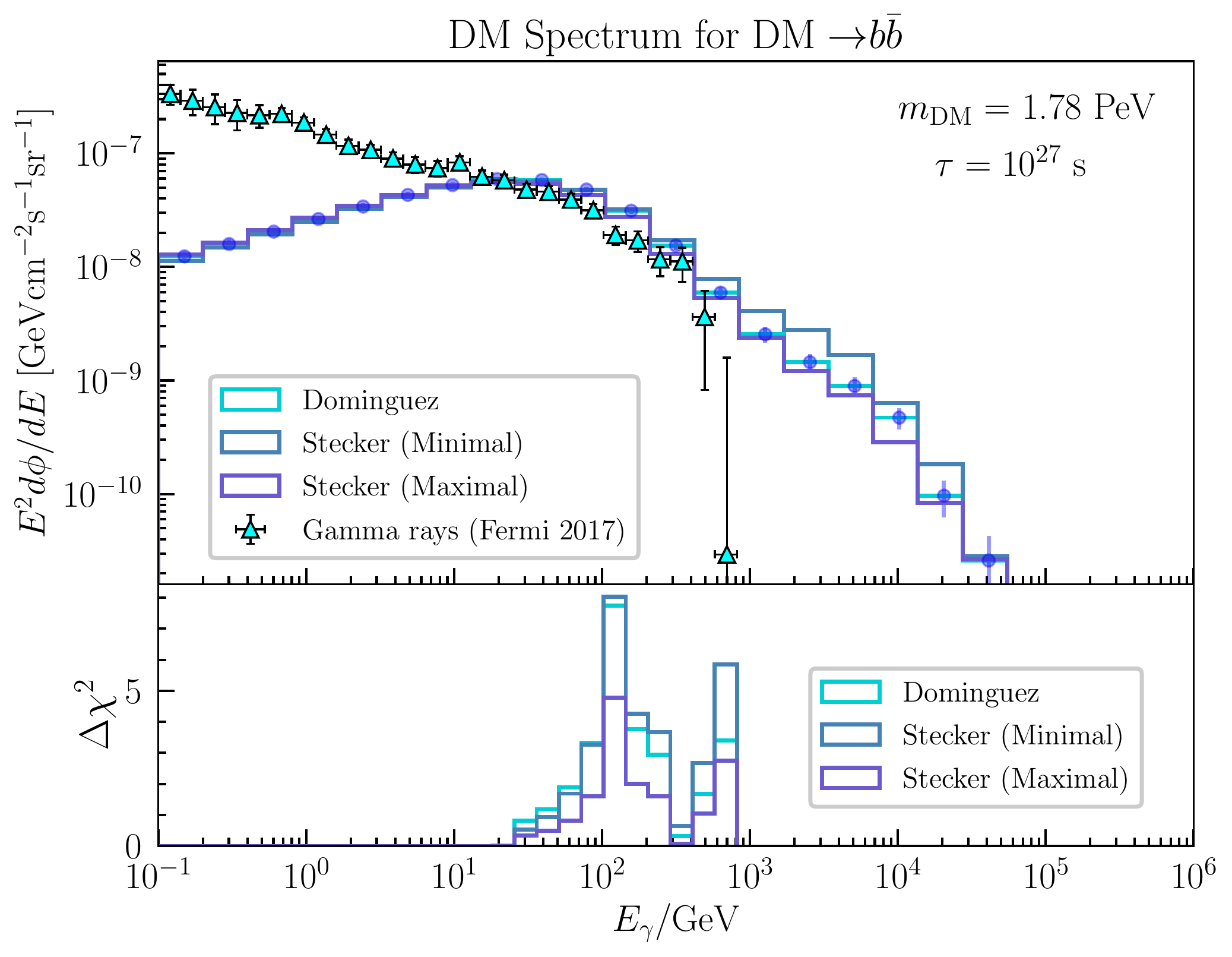}
    \includegraphics[width = 0.24\columnwidth]{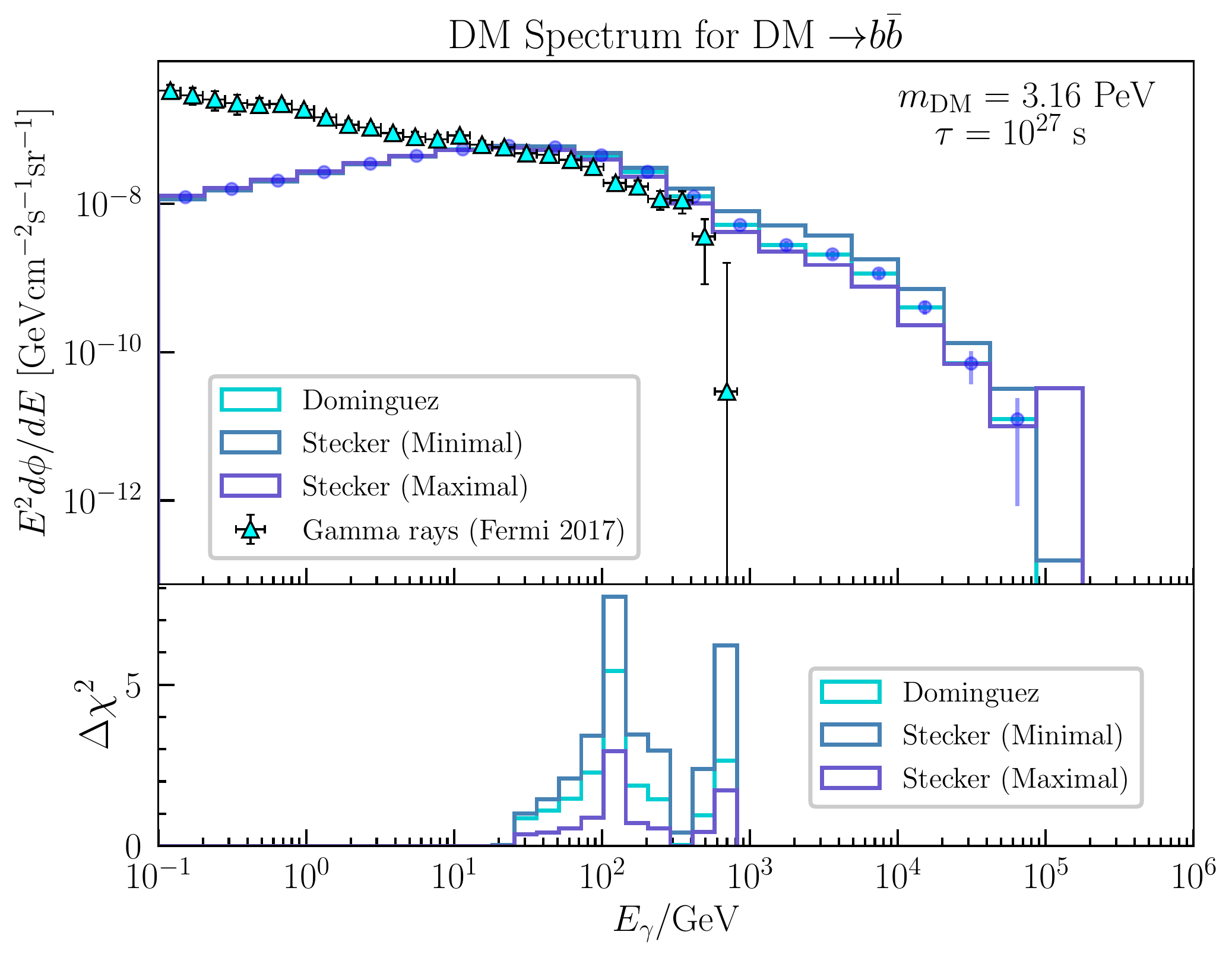}
    \includegraphics[width = 0.24\columnwidth]{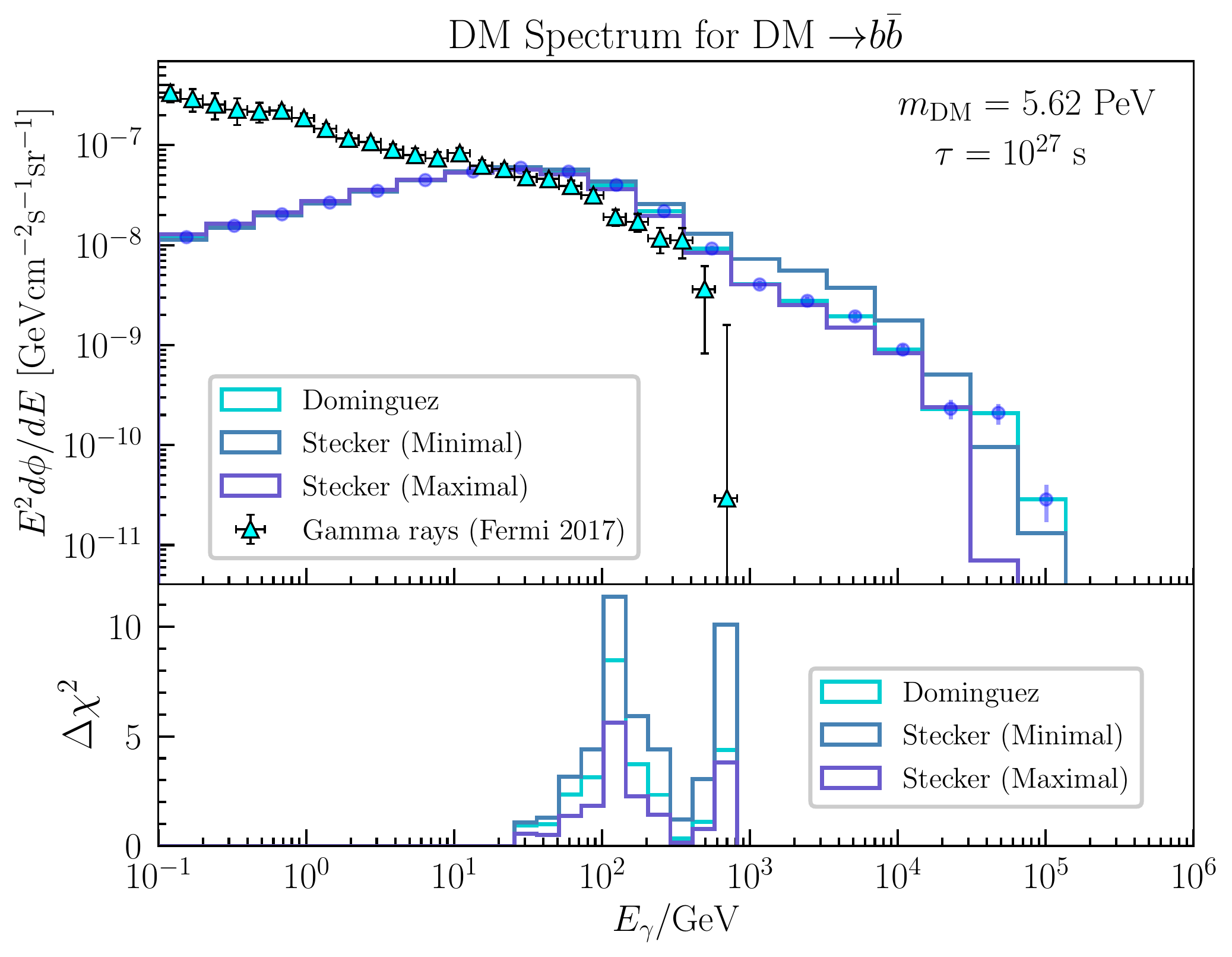}
    \includegraphics[width = 0.24\columnwidth]{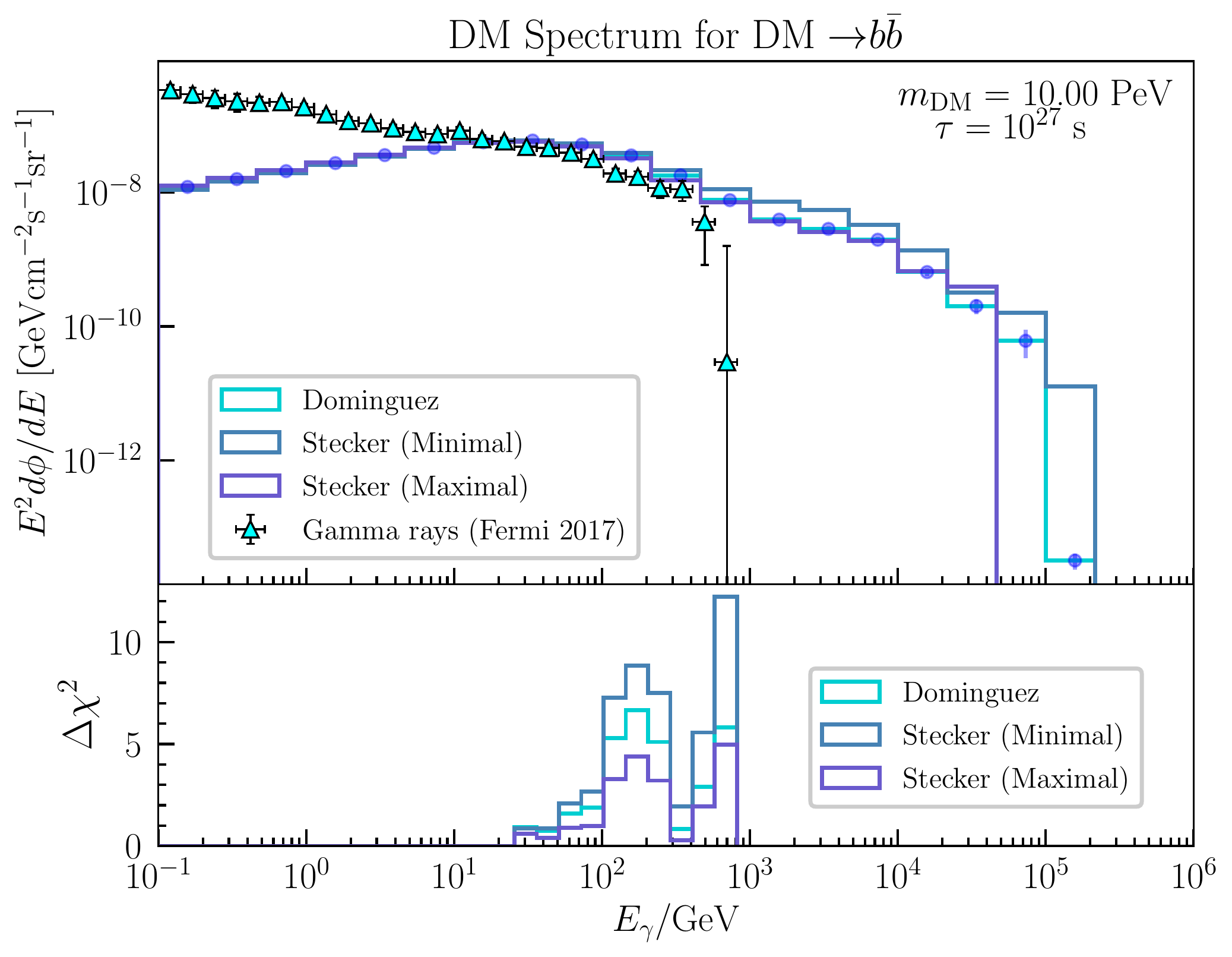}
\end{center}
    \caption{Final dark matter gamma-ray flux from the extragalactic secondary $e^+e^- \to \gamma$ component, for different dark matter masses at a lifetime of $10^{27}$ seconds. Different colors correspond to different EBL models considered for the propagation through the extragalactic space. The bottom panels show the $\Delta \chi^2$ calculated by comparing our Monte Carlo simulations to Fermi-LAT measurements of the IGRB spectrum, displayed by the triangular points. The circular data points show the mean prediction and the corresponding errors from the Monte Carlo simulations according to the Dominguez \textit{et al.} EBL model.}
    \label{fig:EG-Dark-Matter-Electrons}
\end{figure}
\end{landscape}

The different IGRB predictions of the gamma-ray flux for the three EBL models and for different DM masses can be seen in~\cref{fig:EG-Dark-Matter-Gamma} for the $\gamma \to \gamma$ extragalactic secondary component, and in~\cref{fig:EG-Dark-Matter-Electrons} for the $e^+e^-\rightarrow \gamma$ one.

\end{document}